\documentclass[fleqn,usenatbib]{mnras}

\usepackage{newtxtext,newtxmath}

\usepackage[T1]{fontenc}

\DeclareRobustCommand{\VAN}[3]{#2}
\let\VANthebibliography\thebibliography
\def\thebibliography{\DeclareRobustCommand{\VAN}[3]{##3}\VANthebibliography}

\usepackage{graphicx}	
\usepackage{amsmath}	

\title[Superdiffusion of Cosmic Rays in Compressible Magnetized Turbulence]{Superdiffusion of Cosmic Rays in Compressible Magnetized Turbulence}

\author[Hu et al.]{
Yue Hu$^{1,2}$\thanks{E-mail: yue.hu@wisc.edu}
, A. Lazarian$^{2,3}$\thanks{E-mail: alazarian@facstaff.wisc.edu}
, Siyao Xu$^{4}$\thanks{E-mail: sxu@ias.edu (Hubble Fellow)}
\\
$^{1}$Department of Physics, University of Wisconsin-Madison, Madison, WI, 53706, USA\\
$^{2}$Department of Astronomy, University of Wisconsin-Madison, Madison, WI, 53706, USA\\
$^{3}$Centro de Investigación en Astronomía, Universidad Bernardo O’Higgins, Santiago, General Gana 1760, 8370993,
Chile\\
$^{4}$Institute for Advanced Study, 1 Einstein Drive, Princeton, NJ 08540, USA\\
}

\date{Accepted XXX. Received YYY; in original form ZZZ}

\pubyear{2020}

\begin{document}
\label{firstpage}
\pagerange{\pageref{firstpage}--\pageref{lastpage}}
\maketitle

\begin{abstract}
Owing to the complexity of turbulent magnetic fields, modeling the diffusion of cosmic rays is challenging. Based on the current understanding of anisotropic magnetohydrodynamic (MHD) turbulence, we use test particles to examine the cosmic rays' superdiffusion in the direction perpendicular to the mean magnetic field. By changing Alfv\'en Mach number $M_A$ and sonic Mach number $M_S$ of compressible MHD simulations, our study covers a wide range of astrophysical conditions including subsonic warm gas phase and supersonic cold molecular gas. We show that freely streaming cosmic rays' perpendicular displacement increases as 3/2 to the power of the time traveled along local magnetic field lines. This power-law index changes to 3/4 if the parallel propagation is diffusive. We find that the cosmic rays' parallel mean free path decreases in a power-law relation of $M_A^{-2}$ in supersonic turbulence. We investigate the energy fraction of slow, fast, and Alfv\'enic modes and confirm the dominance of Alfv\'enic modes in the perpendicular superdiffusion. In particular, the energy fraction of fast mode, which is the main agent for pitch-angle scattering, increases with $M_A$, but is insensitive to $M_S\ge2$. Accordingly, our results suggest that the suppressed diffusion in supersonic molecular clouds arises primarily due to the variations of $M_A$ instead of $M_S$.
\end{abstract}




\section{Introduction}
\label{sec:intro}

Diffusive propagation of cosmic rays (CRs) is a fundamental topic in the analysis of their origin \citep{1969ApJ...156..445K,1975MNRAS.172..557S,1999ApJ...520..204G,2002cra..book.....S,2019PhRvL.122e1101B,2021MNRAS.502.5821F}. The diffusion also affects various astrophysical processes, e.g., stellar modulation for exoplanets \citep{2016A&A...585A..96T,2020EPSC...14..818R}, heating and ionization in molecular clouds \citep{2016ApJ...824...89S,2011ApJ...740L...4C}, driving galactic winds \citep{2017MNRAS.467..906W,2020MNRAS.493.2817K}, and shock acceleration \citep{1987ApJ...313..842J,1996A&A...314.1010K,2012ApJ...750...87P,2021arXiv211104759X}. In a magnetized medium, as magnetic fields constrain CRs' motion, the propagation parallel and perpendicular to the magnetic fields are very different. 

The CRs' propagation along magnetic field is affected by the fluctuations of turbulent magnetic fields, which introduce pitch-angle scattering and diffusive propagation along magnetic field \citep{1966ApJ...146..480J,1996JGR...101.2511B,2002ApJ...578L.117Q}. In addition, CRs can be reflected by magnetic mirrors \citep{1973ApJ...185..153C}. It was shown in \citet{2021arXiv210608362L} (hereafter LX21) that bouncing among magnetic mirrors results in the diffusive propagation parallel to the local mean magnetic field. This is a new type of CR diffusion that was termed in LX21 "mirror diffusion" (see also \citealt{2021arXiv211008282X}). 

The textbook picture of CRs' propagation considers a stochastic magnetic field \citep{1966ApJ...146..480J,1967ApJ...149..405J,1990JGR....9520673M}. The random magnetic lines result in CRs' trajectories spread across the mean magnetic field, as CRs diffuse in the direction normal to the mean field \citep{1969ApJ...155..777J}, and this effect dominates the propagation in the direction perpendicular to the mean magnetic field. As the propagation of CRs parallel to the magnetic field was known to be diffusive \citep{1966ApJ...146..480J}, incidentally the subsequent application of the two diffusive laws would result in the subdiffusive propagation of CRs perpendicular to magnetic field \citep{2002ApJ...578L.117Q}. The suggestion was shown to be inconsistent with the actual propagation in MHD turbulence \citep{YL08}. In particular, the classical assumption of isotropic turbulence is in contrast with what we have learned about MHD turbulence \citep{GS95,LV99,2000ApJ...539..273C}. The actual connection between CRs' propagation and properties of MHD turbulence was obtained after \citet{LV99} established the superdiffusive behavior of magnetic field lines. Being supported numerically \citep{2006MNRAS.373.1195L,2013ApJ...767L..39B}, this became the foundation for the description of perpendicular superdiffusion of cosmic ray idea as well as the reason why the subdiffusion of cosmic rays is very unlikely \citep{YL08,2014ApJ...784...38L}.

The Alfv\'en motions were typically considered for describing CRs' interaction with turbulence. However, in reality, the properties of Alfv\'enic turbulence are radically different from the isotropic models originally employed. The effect of the MHD turbulence's anisotropy has been investigated extensively in the literature \citep{2000PhRvL..85.4656C,2003A&A...397..777T,YL02,YL04,YL08}. In particular, the anisotropy means that turbulent energy preferentially cascades along the direction perpendicular to local magnetic fields making the eddies corresponding to Alfv\'en motions elongated along the magnetic field. This dramatically decreases the efficiency of resonance scattering if the turbulence is injected on scales much larger than the CR Larmor radius. Moreover, importantly the Alfv\'enic turbulence induces superdiffusive divergence of magnetic field lines \citep{LV99}. If CRs propagate ballistically along the magnetic field, this imprints the superdiffusive behavior on the dispersion of CRs perpendicular to the magnetic field. This has important consequences for both CR propagation and the acceleration \citep{YL08,2014ApJ...784...38L}. The superdiffusion of CRs is the focus of the present paper.

The superdiffusion of magnetic field lines is related to the turbulent wandering of magnetic fields described in \citet{LV99}. The wandering means that the magnetic field lines spread out in the perpendicular direction with the distance along the field lines. As derived in \citet{LV99}, the perpendicular separation between magnetic field lines increases as the distance along the field line to the power $3/2$, i.e. as $x^{3/2}$, while the earlier studies assumed the random walk behavior, i.e $x^{1/2}$. This superdiffusive law was confirmed in several numerical studies \citep{2004ApJ...603..180L,2013ApJ...767L..39B,2013PhRvD..87d3008C}. In this case, the perpendicular diffusion of CRs increases $t^{3/2}$, where $t$ is the time moved along the magnetic field.

Considering the situation of freely streaming CRs, i.e., in the absence of scattering, CRs strictly move along the magnetic field lines. CRs' perpendicular displacement is governed by the lines' separation, resulting in a superdiffusion relation \citep{YL08,2014ApJ...784...38L,XY13}. In the presence of efficient scattering, the parallel diffusion slows down the motion of CRs along the diverging magnetic field lines. In this situation, the perpendicular diffusion of CRs increases as $t^{3/4}$ which is still superdiffusion, although a weaker one \citep{YL08,2014ApJ...784...38L}. 

MHD turbulence has an important role in regulating CRs' diffusion. However, the properties of astrophysical turbulence are different in different gas phases and astrophysical media, e.g., highly compressible turbulence in star-forming regions and supernova remnants \citep{2015ApJS..216...18N,2016ApJ...832..143F,2020ApJ...905..159X,IRAM,2021arXiv210913670L,CMZ}, weakly compressible turbulence in the diffuse ISM \citep{1995ApJ...443..209A,2010ApJ...710..853C,2018ApJ...865...46L}. It is important that compressible MHD turbulence consists of slow, fast, and Alfv\'en modes \citep{GS95,2000ApJ...539..273C,2001ApJ...562..279L,2002ApJ...564..291C,2003MNRAS.345..325C}. The energy fractions of Alfv\'en modes and slow and fast modes can be different in the multi-phase ISM. The fast mode is identified as the most efficient agent for scattering in the ISM \citep{YL02,YL04}, which affects the diffusion process. This result persists in the CRs nonlinear scattering theory \citep{YL08}. This calls for a detailed study of CR propagation for different levels and compressibility of MHD turbulence. An appropriate description of turbulent magnetic fields is, therefore, indispensable for modeling the CRs' propagation in magnetized interstellar turbulence. 

This work investigates the CRs' diffusion using realistic turbulent magnetic fields produced by three-dimensional MHD turbulence simulations. The analysis focuses on high-energy CRs, for which the CRs' feedback is negligible, including both subsonic and supersonic regimes covering the majority of astrophysical gas phases. The subsonic case has been numerically tested by \citet{XY13}. In order to obtain sufficient scattering, artificial resonant slab fluctuations were introduced to the initial turbulent magnetic fields in some cases considered there. The resonant slab fluctuations are not considered in our analysis and the scattering comes from the intrinsic properties of MHD simulations. Also, we decompose the MHD turbulence into slow, fast, and Alfv\'en modes and investigate the contribution from each mode to the superdiffusion. 
Our work is also complementary to the recent study in \citet{2021arXiv210801936M}, which 
is similar to the study by \citet{XY13}, but performed turbulence decomposition. We further consider a wide range of turbulence environments, i.e., different sonic Mach number which varies significantly in the multi-phase ISM. For example, the warm ionized medium is usually subsonic \citep{2011Natur.478..214G}, while cold molecular gas appears supersonic \citep{IRAM}. Studying how the fractions of the three modes depend on the sonic Mach number is therefore important in understanding CRs' propagation. 

The paper is organized as follows. In \S~\ref{sec:theory}, we derive the perpendicular superdiffusion of CRs induced by diverging magnetic field lines. In \S~\ref{sec:data}, we provide the details of the numerical simulations used in this work. \S~\ref{sec:method}, we illustrate the recipe of tracing CRs' trajectory by post-processing the MHD simulations. In \S~\ref{sec:results}, we present the numerical testing of CRs' perpendicular superdiffusion, pitch-angle scattering, and MHD mode decomposition followed by discussions and conclusion in \S~\ref{sec.dis} and \S~\ref{sec:con}, respectively.

\section{Theoretical consideration}
\label{sec:theory}
\subsection{Anisotropy of Aflv\'enic turbuluence}
To understand the process of magnetic line separation in a turbulent medium, i.e., wandering of magnetic fields, it is necessary to consider the anisotropy of MHD turbulence. The scaling for incompressible turbulence was proposed in GS95 for the setting of turbulence having velocity $v_{\rm inj}$ equal to the Alfv\'en velocity $v_A$ at the injection scale. This setting corresponds to the Alfv\'en Mach number $M_A=v_{\rm inj}/v_A$ equal to unity. The anisotropy of turbulence in GS95 was given by the relation
\begin{equation}
\label{eq.gs95}
    k_\parallel\propto k_\bot^{2/3}
\end{equation}
where $k_\parallel$ and $k_\bot$ are wavevectors parallel and perpendicular to the magnetic field. Originally, in GS95 these vectors were defined with respect to the mean magnetic field, i.e. in the global system of reference, which is still a frequent confusion among CRs researchers who try to use the GS95 relation. In fact, the theory of turbulent reconnection in LV99 presents a different outlook on the nature of MHD motions. It suggests that the magnetic field presents a collection of eddies that mix up magnetic fields perpendicular to the direction of their local ambient magnetic field.  In this setting, the mean global magnetic field gets irrelevant, while the  "critical balance" condition introduced by GS95 has a very simple physical interpretation (i.e. the eddy turnover time $l_\bot /v_l$ equals the wave period $l_\parallel/ v_A$) that is induced by the eddy motion. Above, $v_l$ is turbulent velocity at scale $l$ and $v_A$ is Alfv\'{e}n speed. This type of dynamics is possible as the turbulent reconnection reconnects the magnetic field of the eddy within one eddy turnover time and thus it is easier to mix magnetic field lines in the direction perpendicular to the magnetic field rather than bend the field lines. In other words, in strong MHD turbulence, the energy is channeled along the path of the minimal resistance along the perpendicular direction. Naturally, in this case, the spectrum of hydro motions is Kolmogorov, i.e., $v_l\propto l_\bot^{1/3}$, where $l_\bot$ is the perpendicular scale of the turbulent eddy. Incidentally, this scaling of the turbulence anisotropy is not observable in the global reference frame defined by mean magnetic fields, due to the averaging effect \citep{2000ApJ...539..273C}. At the same time, the CRs sample the magnetic field locally and experience the aforementioned anisotropy. 

The importance of the local system of reference was demonstrated numerically in a seminal paper by \citep{2000ApJ...539..273C}.  The turbulence scaling for sub-Alfv\'enic turbulence  was introduced in \citet{LV99}:
\begin{equation}
\label{eq.lv99}
 l_\parallel= L_{\rm inj}(\frac{l_\bot}{L_{\rm inj}})^{\frac{2}{3}}{M_A^{-4/3}}, ~~~~{ M_A\le 1}
\end{equation} 
where $L_{\rm inj}$ is the turbulence injection scale. 

\subsection{CRs perpendicular propagation induced by wandering magnetic field lines}
This anisotropy (see Eq.~\ref{eq.lv99}) is crucial as it gives an insight into how the separation of magnetic field lines grows when moving along the magnetic field lines. 

Following \citet{2014ApJ...784...38L}, we firstly consider the case that the parallel mean free path $\langle\lambda_\parallel\rangle$ of CRs is much larger than the injection scale of turbulence. Consequently, the scattering of CRs is negligible, and their diffusion perpendicular to the mean magnetic field is determined by the divergence of magnetic field lines. 

\begin{figure}
	\centering
	\includegraphics[width=1.0\linewidth]{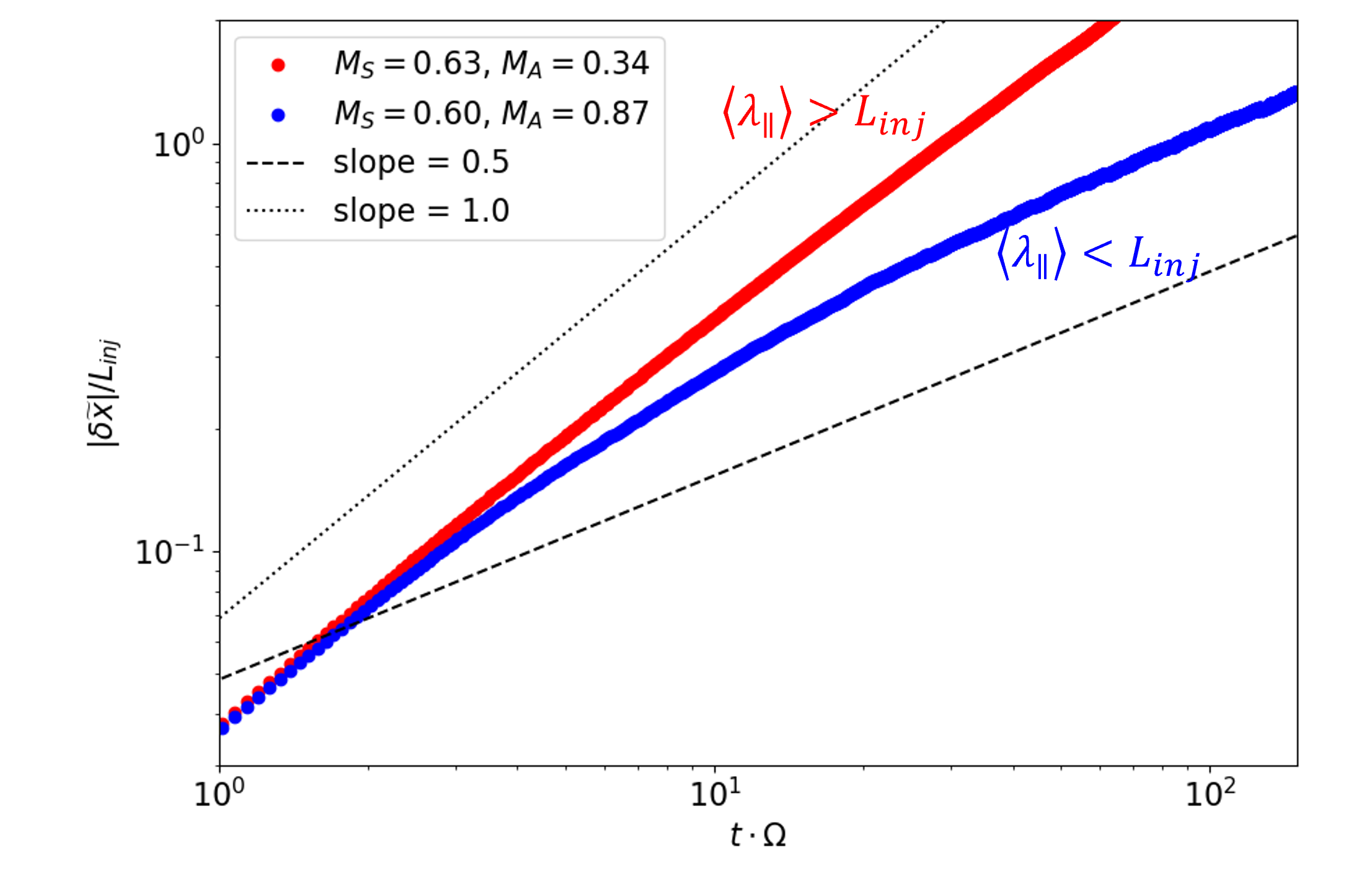}
	\caption{\label{fig:Dx} An example of CRs' parallel transport $|\delta \widetilde{x}|= \sqrt{\langle(x-x_0)^2\rangle}$ along the magnetic field, where $x_0$ is the initial position of particles. $k$ denotes the power-law slope of the reference lines.}
\end{figure}

Supposing that when CRs moves a distance $x$ in the direction parallel to local magnetic field lines, the field lines spread out in the perpendicular direction with distance $l_\bot$. Consequently, one can express the rate of field line diffusion as:
\begin{equation}
    \frac{d\langle z^2\rangle}{dx}\approx\frac{d}{dx}l_\bot^2\approx\frac{l_\bot^2}{l_\parallel}
\end{equation}
where $\sqrt{\langle z^2\rangle}$ is the ensemble averaged diffusion perpendicular to magnetic field lines. Considering the anisotropy relation given by Eq.~\ref{eq.lv99}, we get:
\begin{equation}
\label{eq.4}
\frac{d}{dx}l_\bot^2\approx L_{\rm inj}(\frac{l_\bot}{L_{\rm inj}})^{\frac{4}{3}}{M_A^{4/3}},~~{ M_A\le 1}
\end{equation}
The solution of Eq.~\ref{eq.4} is
\begin{equation}
\label{eq.5}
l_\bot^2 \approx \frac{x^3}{27L_{\rm inj}}M_A^{4},~~{ M_A\le 1}
\end{equation}
which indicates that the diffusion $\sqrt{\langle z^2\rangle}$ of magnetic field lines increases as $x^{3/2}$. Owing to the fact the scattering is negligible here, CRs are freely streaming along magnetic field lines with the velocity $c\mu$, where $c$ is the light speed and $\mu$ is the pitch angle cosine so that $x\sim c\mu t$ (see Fig.~\ref{fig:Dx}), here $t$ is time. Consequently, we have the superdiffusion in the direction perpendicular to magnetic field:
\begin{equation}
\label{eq.6}
    \langle z^2\rangle\approx l_\bot^2 \propto \frac{t^3}{27L_{\rm inj}}M_A^{4},~~{ M_A\le 1}
\end{equation}
Note, this is valid only for strong sub-Alfv\'enic turbulence, i.e., for perpendicular scales that are smaller than $l_{\rm tr}=L_{\rm inj}M_A^2$. 
Magnetic field wandering is the source of perpendicular superdiffusion of CRs as presented in Eq.~\ref{eq.6}. As was established in LV99 the perpendicular squared displacement of the magnetic field is proportional to the cube of the path of the displacement along the magnetic field, i.e. $x^3$, which resembles the relation of the squared separation of particles with cubic of time, i.e. $t^3$ present in the hydrodynamic turbulence. The latter is known as the Richardson diffusion \citep{1926RSPSA.110..709R}. The close relation between the LV99 dispersion of magnetic field lines and the Richardson diffusion in a magnetized fluid is established in \citet{2013Natur.497..466E}. Physically, the analogy between the hydrodynamic behavior and that of MHD arises from the fast magnetic reconnection in turbulent fluids that allows free eddy motions in the direction perpendicular to the local direction of the magnetic field \citep{2020PhPl...27a2305L}. The superdiffusion of CRs in MHD turbulence arises from the Richardson diffusion of magnetic field lines in terms of magnetic field wandering \citep{2014ApJ...784...38L}.

Weak sub-Alfv\'enic turbulence (i.e., $l_\bot>l_{\rm tr}$) is wave-like and requires a different treatment \citep{LV99}:
\begin{equation}
    l_\bot^2 \sim xL_{\rm inj} M_A^4,
\end{equation}
which reflects the usual diffision behavior. The special feature of this diffusion is that it takes place on scales $x>L_{\rm inj}$, but the step of the diffusion is $M_A L_{\rm inj}<L_{\rm inj}$. This reflects the anisotropy of turbulence on the injection scale predicted in \citep{LV99}.

When CRs move along magnetic field lines in a non-ballistic way and have mean free path $\langle\lambda_\parallel\rangle$ less than $L_{\rm inj}$, the propagation along magnetic field lines is diffusive. This diffusion is imprinted, as discussed in \citet{2014ApJ...784...38L}, in the CRs' diffusion both in the directions parallel and perpendicular to the local mean magnetic field. However, the mean angle of the magnetic field lines is changing with $x$ and therefore we have parallel diffusion $\langle x^2 \rangle$ and perpendicular diffusion $\langle z^2 \rangle$:
\begin{equation}
\begin{aligned}
        \delta\langle x^2 \rangle &\approx D_\| \delta t\\
        \delta\langle z^2 \rangle &\approx D_\bot \delta t
\end{aligned}
\end{equation}
where $D_\|$ and $D_\bot$ are parallel and perpendicular coefficients, respectively (see Fig.~\ref{fig:Dx}). Considering that the projection of the mean free path $\langle\lambda_\parallel\rangle$to the perpendicular direction is approximately $\langle\lambda_\parallel\rangle l_\bot/x$, one can get $D_\bot\approx D_\| l_\bot/x$. Consequently, we have:
\begin{equation}
\label{eq.9}
    \langle z^2\rangle \approx D_\bot t\approx l_\bot/x D_\| t\propto D_\|t^{3/2}
\end{equation}
where $l_\bot/x\propto t^{1/2}$ came from magnetic field wondering (see Eq.~\ref{eq.6}\footnote{Eq.~\ref{eq.6} and Eq.~\ref{eq.9} can expressed in terms of the distance parallel to local magnetic field lines: 
\begin{equation}
\begin{aligned}
        \langle z^2\rangle&\propto M_A^{4}x^3,~~{ M_A\le 1}\\
        \langle z^2\rangle&\propto D_\|x^{5/2},~~{ M_A> 1}
\end{aligned}
\end{equation}}). As demonstrated in \citet{2014ApJ...784...38L} , the diffusion of CRs in respect to the local and global system of reference can differ. This was confirmed by a recent numerical study in \citet{2021arXiv210801936M}. Anticipating the approximate nature of the scaling employed, we do not attempt to present the exact numerical factor for the dependence.

\begin{figure}
	\centering
	\includegraphics[width=1.0\linewidth]{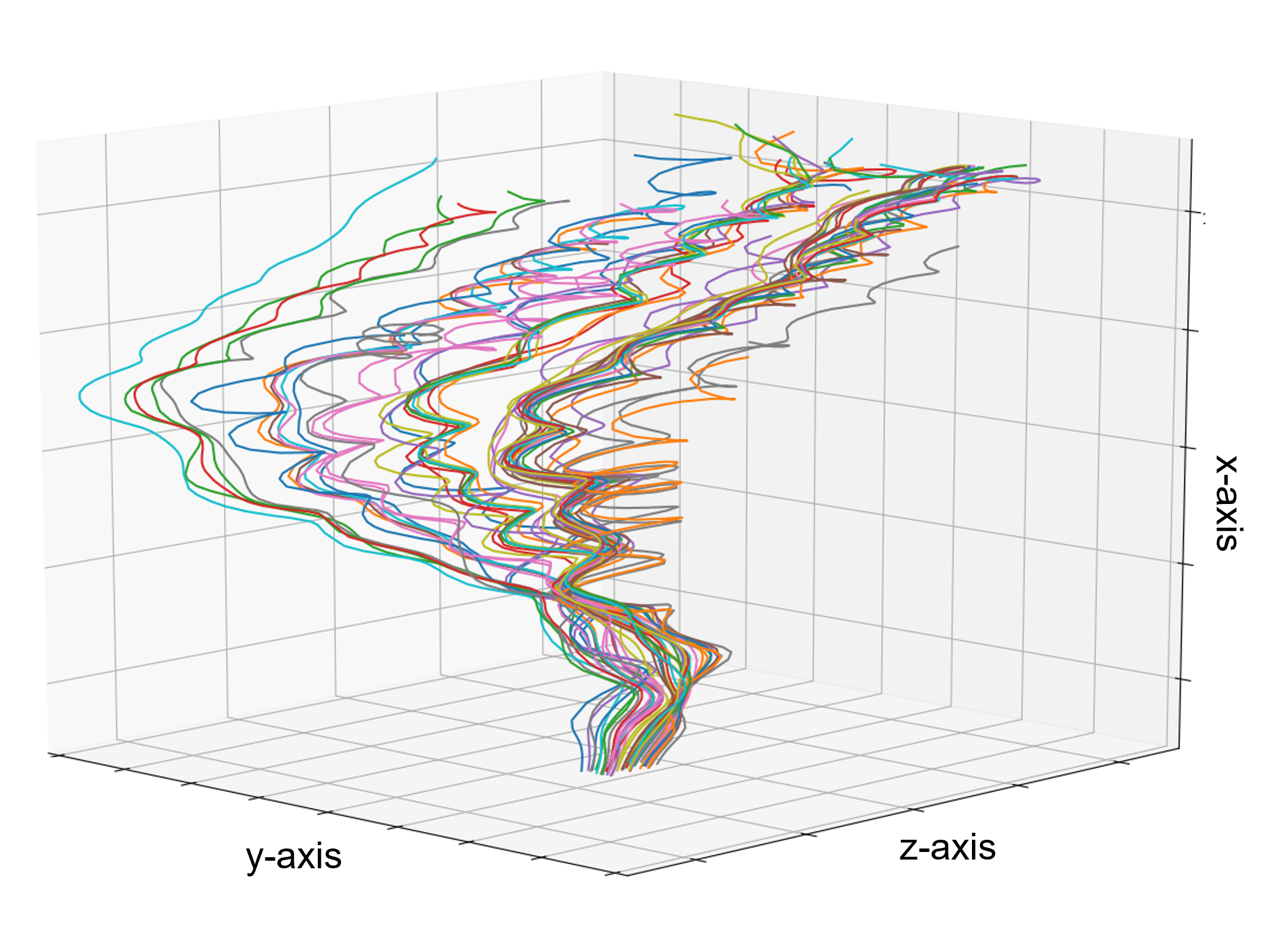}
	\caption{\label{fig:exmaple} An visualization of CRs' superdiffusion. Fifty CRs' trajectories are shown with $M_S=0.62$ and $M_A=0.56$. The Larmor radius $r_L$ is set as 1/100 of cube size $L$. The spatial separation between CRs is one pixel and initial pitch angle is 0 degree. }
\end{figure}

In super-Alfv\'enic condition, turbulence is hydro-like in the range of scales [$l_A$, $L_{\rm inj}$] because the dynamics of magnetic fields is dominated by turbulent motions for which the magnetic back reaction is negligible. Above, $l_A=L_{\rm inj}M_A^{-3}$, is the scale at which the turbulent velocity equals the Alfv\'en speed. 

The role of magnetic field becomes more important at scales smaller than $l_A$ but larger than the dissipation scale \citep{2006ApJ...645L..25L}. In this scale range, turbulent eddy gets elongated in the direction of the local magnetic field in a way similar to the sub-Alfv\'enic turbulence and $l_A$ plays the role of injection scale: $v_l\propto v_A(l_\bot/l_A)^{1/3}$. Consequently, the anisotropy relation in ``critical balance" is:
\begin{equation}
\label{eq.l06}
 l_\parallel= L_{\rm inj}(\frac{l_\bot}{L_{\rm inj}})^{\frac{2}{3}}{M_A^{-1}},~~{ M_A\ge 1}
\end{equation} 
and we have:
\begin{equation}
\langle z^2\rangle \approx l_\bot^2 = \frac{x^3}{27L_{\rm inj}}M_A^{3}\propto \frac{t^3}{27L_{\rm inj}}M_A^{3},~~{ M_A\ge 1}
\end{equation}

The above relations can be obtained formally considering that in the case of super-Alfv\'enic turbulence the injection happens scale is $l_A$ rather than $L_{\rm inj}$. Therefore, on the scales from the dissipation scale to $l_A$ one can use the relations and all what we discussed above in terms of CR propagation for sub-Alfv\'enic turbulence is applicable with this change. 

It is important for the CR propagation that the perturbations of magnetic fields in a turbulent medium can scatter and isotropize cosmic rays. This is the source of the generally accepted parallel to magnetic field diffusion. This is not the only process that governs CR parallel diffusion, however. A new process of mirror diffusion was introduced in \citet{2021arXiv210608362L}. Unlike the scattering, this diffusion does not cause the diffusion in CR pitch angles\footnote{\citet{2021arXiv210608362L} pointed out that the dynamics of the pitch angle similar to the pitch angle dynamics for the Transient Time Damping (TTD) acceleration discussed e.g. in \citet{2006A&A...454....1S}.}. Instead CRs bounce from the magnetic mirrors created by magnetic compression. While the earlier studies (see \citealt{1969ApJ...156..445K}) assumed that the magnetic mirrors will induce trapping of CRs, \citet{2021arXiv210608362L} demonstrated that due to magnetic field wondering that we discussed above, the CRs were not trapped, but every time bounce from a new mirror and, as a result, diffuse parallel to magnetic field. This new type of diffusion, i.e. the "mirror diffusion" is particularly important for the CRs propagation new CR sources \citep{2021arXiv211008282X}. However, for the sake of simplicity, within the present study, we do not distinguish between the two types of parallel to magnetic field diffusion. 

\subsection{Compressible MHD turbulence}

Compressible turbulence is a non-linear phenomenon, which nevertheless allows an approximate representation as a composition of three cascades (see \citealt{BL19} and ref. therein). The Alfv\'en cascade imposes its properties on slow mode turbulence cascade, while the fast mode cascade evolves mostly on its own \citep{GS95,2001ApJ...562..279L,2002ApJ...564..291C,2003MNRAS.345..325C}. The exchange of energy between different modes was proven numerically to be insignificant \citep{2002PhRvL..88x5001C} and this substantially simplifies the practical treatment of MHD turbulence. \citet{LV99} showed that Alfv\'enic turbulence induces magnetic field wandering. Therefore, the CRs' superdiffusion is dominated by Alfv\'enic modes of turbulence. Fig.~\ref{fig:exmaple} presents an example of CRs' superdiffusion using the method explained in \S~\ref{sec:method}. Initially the spatial separation between CRs is one pixel in the direction perpendicular to the mean magnetic field field. The separation, however, explosively grows up when CRs travel along the magnetic field. However, fast modes dominate the scattering, which can affect the scaling of superdiffusion (see Eq.~\ref{eq.6}). Therefore we will investigate superdiffusion in compressible MHD turbulence.

\begin{table}
	\centering
	\label{tab:sim}
	\begin{tabular}{| c | c | c | c | c |}
		\hline
		Model & $M_S$ & $M_A$ & Resolution & $\beta$ \\	\hline\hline
		A0 & 0.63 & 0.34 & $792^3$ & 0.58 \\
		A1 & 0.62 & 0.56 & $792^3$ & 1.63 \\
		A2 & 0.60 & 0.78 & $792^3$ & 3.38 \\
		A3 & 0.60 & 0.87 & $792^3$ & 4.21 \\\hline
		B0 & 0.14 & 0.65 & $792^3$ & 43.11 \\
		B1 & 0.39 & 0.60 & $792^3$ & 4.73 \\
		B2 & 0.84 & 0.54 & $792^3$ & 0.83 \\
		B3 & 1.06 & 0.52 & $792^3$ & 0.54 \\
		B4 & 1.27 & 0.50 & $792^3$ & 0.31 \\\hline
		C0 & 7.31 & 0.22 & $792^3$ & 0.01 \\
		C1 & 6.10 & 0.42 & $792^3$ & 0.01 \\
		C2 & 6.47 & 0.61 & $792^3$ & 0.02 \\
		C3 & 6.14 & 0.82 & $792^3$ & 0.04 \\
		C4 & 6.03 & 1.01 & $792^3$ & 0.06 \\
		C5 & 6.08 & 1.19 & $792^3$ & 0.08 \\
		C6 & 6.24 & 1.38 & $792^3$ & 0.10 \\
		C7 & 5.94 & 1.55 & $792^3$ & 0.14 \\
		C8 & 5.80 & 1.67 & $792^3$ & 0.17 \\
		C9 & 5.55 & 1.71 & $792^3$ & 0.19 \\\hline
		D0 & 2.17 & 0.51 & $512^3$ & 0.110 \\
		D1 & 4.35 & 0.52 & $512^3$ & 0.030 \\
		D2 & 6.53 & 0.52 & $512^3$ & 0.010 \\
		D3 & 8.55 & 0.51 & $512^3$ & 0.007 \\		\hline
	\end{tabular}
	\caption{Description of MHD simulations. The compressibility of turbulence is characterized by $\beta=2(\frac{M_A}{M_S})^2$.}
\end{table}
\begin{figure*}
	\centering
	\includegraphics[width=.99\linewidth]{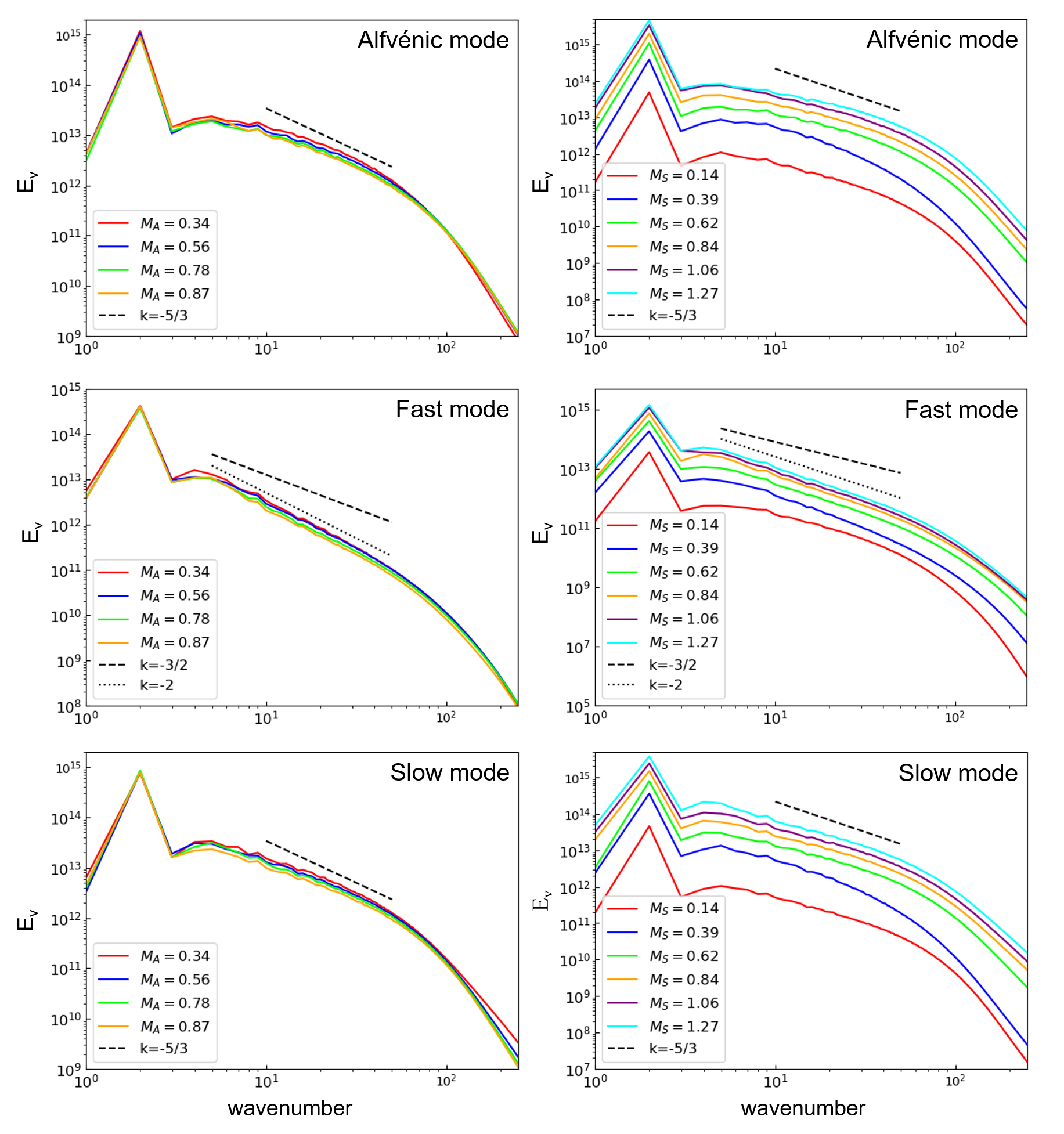}
	\caption{\label{fig:vspectrum} \textbf{Left:} The energy spectrum of velocity fluctuations for Alfv\'enic (top), fast (central), and slow (bottom) modes, in the condition of $M_S\approx0.6$. \textbf{Right:} The energy spectrum of velocity fluctuations for Alfv\'enic, fast, and slow mode, with the condition of $M_A\approx0.5$. k denotes the slope of reference lines.}
\end{figure*}

\section{MHD turbulence simulations}
\label{sec:data}
The 3D compressible MHD turbulence simulations are generated through ZEUS-MP/3D code \citep{2006ApJS..165..188H}. By considering isothermal single fluid and operator-split MHD conditions in the Eulerian frame, we solve the ideal MHD equations in a periodic box:
\begin{equation}
    \begin{aligned}
      &\partial\rho/\partial t +\nabla\cdot(\rho\pmb{v})=0\\
      &\partial(\rho\pmb{v})/\partial t+\nabla\cdot[\rho\pmb{v}\pmb{v}+(P+\frac{B^2}{8\pi})\pmb{I}-\frac{\pmb{B}\pmb{B}}{4\pi}]=\pmb{f}\\
      &\partial\pmb{B}/\partial t-\nabla\times(\pmb{v}\times\pmb{B})=0\\
      &\nabla \cdot\pmb{B}=0
    \end{aligned}
\end{equation}
where $\pmb{f}$ is a random large-scale driving force, $\rho$ is gas density, $\pmb{v}$ is velocity, and $\pmb{B}$ is the magnetic field. We consider the magnetic field and density field in the form of $\boldsymbol{B}=\boldsymbol{B}_0+\delta\boldsymbol{b}$ and $\rho=\rho_0+\delta\rho$, where $\boldsymbol{B}_0$ and $\rho_0$ are the uniform background field. $\delta\boldsymbol{b}$ and $\delta\rho$ stand for fluctuations. Initially, $\boldsymbol{B}_0$ is assumed to be parallel to the x-axis. The gas pressure is given by the isothermal equation of state $P = c_s^2\rho_0$, where $c_s$ is the sound speed. 

We solenoidally (i.e., divergence-free) drive turbulence in Fourier space by applying a stochastic forcing to the momentum equation. In the absence of self-gravity, the forcing is constructed so that kinetic energy is continuously injected on scales that correspond to wavenumbers $k=2$. The driving amplitude is largest at $k=2$ and drops to zero on either side of $k=2$, as shown in Fig.~\ref{fig:vspectrum}. We run the simulation until turbulence is fully developed and the spectrum follows a Kolmogorov scaling. The simulation is grid into 792$^3$ or 512$^3$ pixels and turbulence gets numerically dissipated at scales $\approx$ 10 - 20 pixels.

The scale-free MHD simulations are characterized by the sonic Mach number $M_S=v_{\rm inj}/c_{s}$ and Alfv\'{e}nic Mach number $M_A=v_{\rm inj}/v_{A}$,  where $c_s$ is the isothermal sound speed, $v_{\rm inj}$ is the turbulent velocity at injection scale, and $v_{A}=B_0/\sqrt{4\pi\rho_0}$ is the Alfv\'{e}nic velocity. By varying the values of the background magnetic field and density field, we achieve various $M_A$ and $M_S$ values. The injected level of total turbulent energy is controlled by $M_S$. Several simulations have been used in \cite{YL17b} and \citet{H2}. In this text, we refer to the simulations in Tab.~\ref{tab:sim} by their model name or parameters.
\begin{figure*}[htb!]
	\centering
	\includegraphics[width=1.0\linewidth]{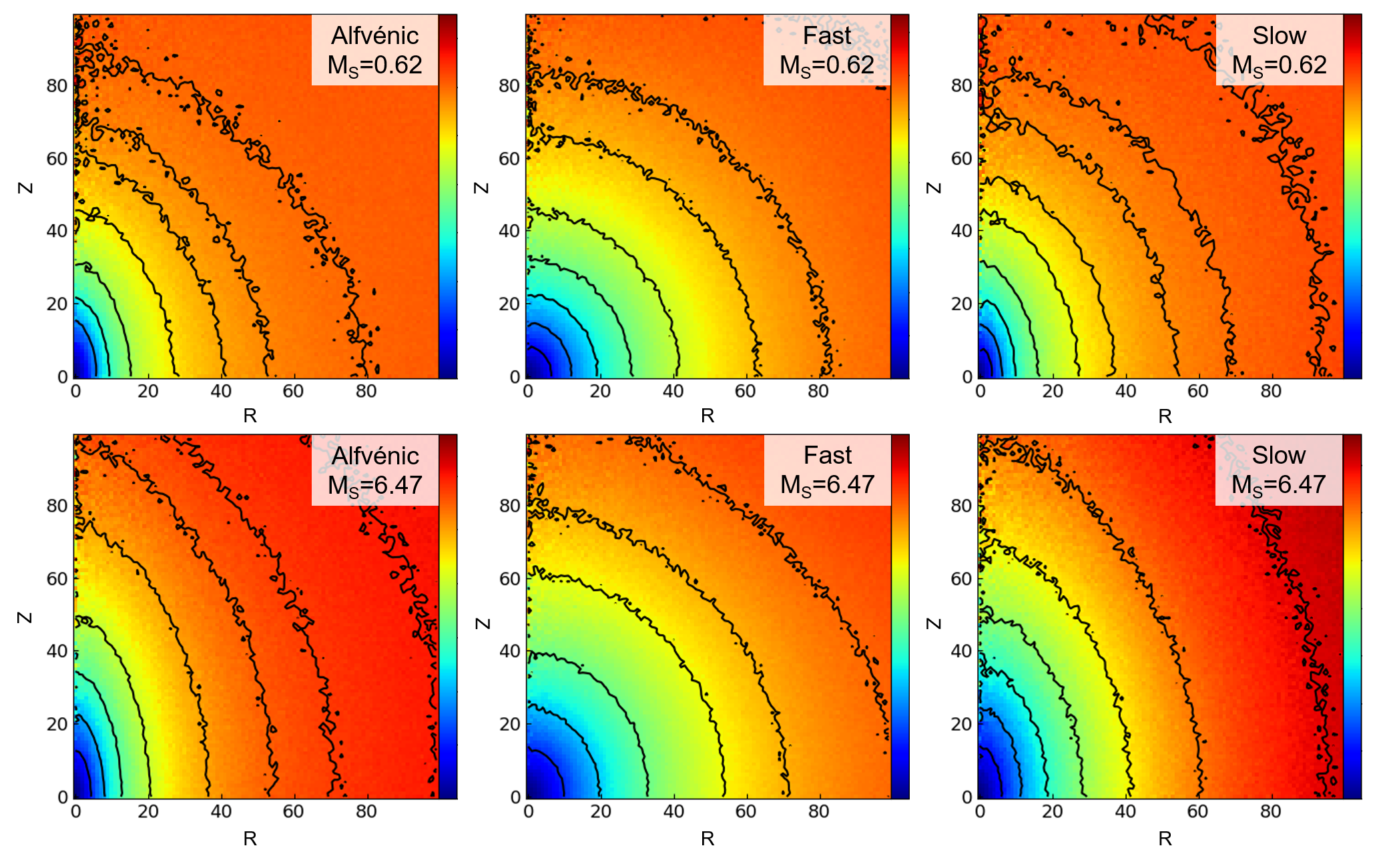}
	\caption{\label{fig:SF_map} Two examples of structure function $SF(R,Z)$ in the local reference frame. $Z$ is parallel to the magnetic field while $R$ is perpendicular to the magnetic field. The two simulations used here both have $M_A\approx0.5$.}
\end{figure*}

\section{Methodology}
\label{sec:method}
\subsection{Tracing particle's trajectory}
To trace the trajectories of CRs, we use the method developed in \cite{2011ApJ...728...60B} and \cite{XY13}. Magnetic fields are extracted directly from MHD turbulence simulations described above. Considering the fact that the relativistic particles have speed much higher than the plasma's Alfv\'en speed, we treat the magnetic field as stationary and neglect the effect of electric field. At each position of a test particle, we compute the Lorentz force on each particle:
\begin{equation}
    \frac{d\boldsymbol{v}}{dt}=\frac{q}{mc}\boldsymbol{v}\times\boldsymbol{B}
\end{equation}
where $\boldsymbol{v}$ is the particle’s velocity, $q$ is particle's charge, $m$ stands for the relativistic mass, $c$ is the light speed, and $\boldsymbol{B}$ is  the local magnetic field. The trajectory is then obtained by integrating the Lorentz force.

The integration employs the Bulirsch-Stoer method with adaptive time step \citep{1986nras.book.....P}. In the case of that the position of a test particle does not locate at integer grid, the corresponding magnetic fields are interpolated using a cubic spline routine. We also adopt periodic boundary conditions.
\subsection{Decomposition of Alfv\'en, fast, and slow modes}
To investigate the effect of different MHD turbulence modes, we employ the mode decomposition method proposed in \citet{2003MNRAS.345..325C}. The decomposition is performed in Fourier space by projecting the velocity's Fourier components on the direction of the displacement vectors $\hat{\boldsymbol{\zeta_A}}$, $\hat{\boldsymbol{\zeta_f}}$, and $\hat{\boldsymbol{\zeta_s}}$ for Alfv\'en, fast, and slow modes, respectively. The displacement vectors are defined as:
\begin{equation}
    \begin{aligned}
    \hat{\boldsymbol{\zeta_A}}&\propto\hat{\boldsymbol{k}}_\bot\times\hat{\boldsymbol{k}}_\parallel\\
    \hat{\boldsymbol{\zeta_f}}&\propto(1+\beta/2+\sqrt{d})k_\bot\hat{\boldsymbol{k}}_\bot+(-1+\beta/2+\sqrt{d})k_\parallel\hat{\boldsymbol{k}}_\parallel\\
    \hat{\boldsymbol{\zeta_s}}&\propto(1+\beta/2-\sqrt{d})k_\bot\hat{\boldsymbol{k}}_\bot+(-1+\beta/2-\sqrt{d})k_\parallel\hat{\boldsymbol{k}}_\parallel\\
    \end{aligned}
\end{equation}
where wavevectors $\boldsymbol{k}_\parallel$ and $\boldsymbol{k}_\bot$ are the parallel and perpendicular to the mean magnetic field $\boldsymbol{B}_0$, respectively. $d = (1 + \beta/2)^2 - 2\beta\cos^2 \vartheta$, $\beta=2(M_A/M_S)^2$, and $\cos \vartheta=\hat{\boldsymbol{k}}_\parallel\cdot\hat{B}_0$. 
\begin{figure*}
	\centering
	\includegraphics[width=1.0\linewidth]{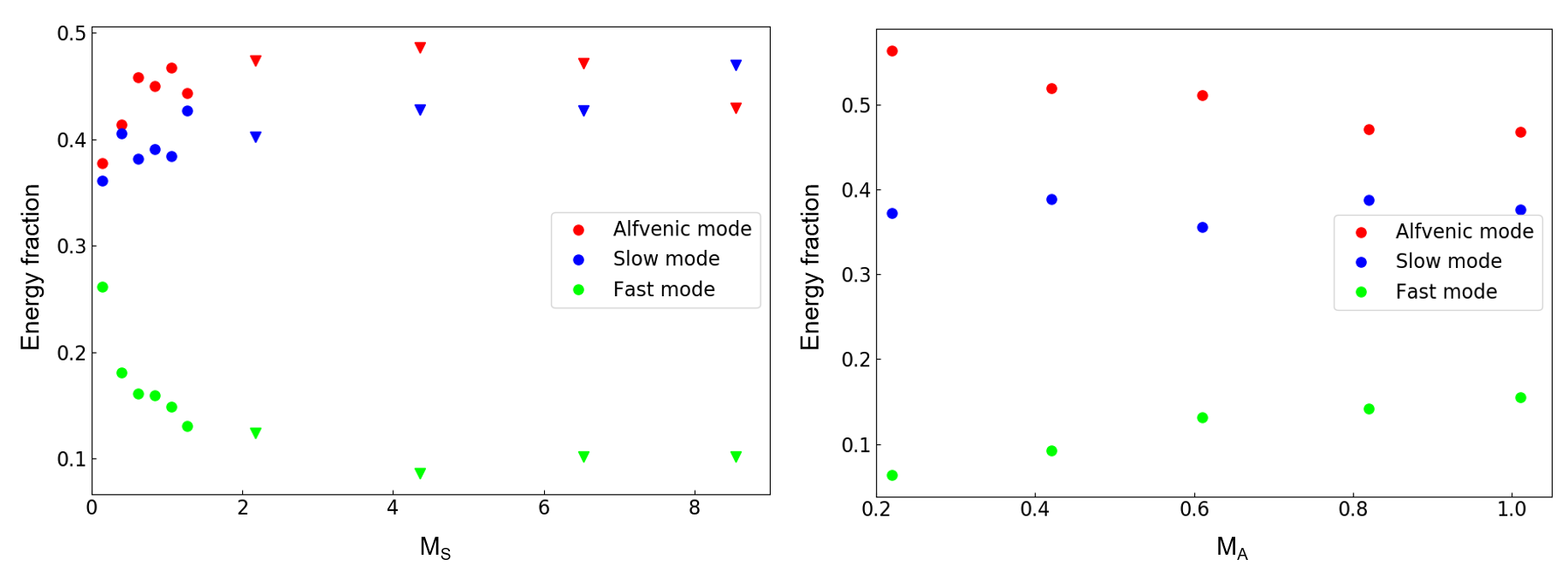}
	\caption{\label{fig:mode} \textbf{Left:} The energy fraction for fast, slow, and Alfv\'enic mode as a function of $M_S$. All simulations have $M_A\approx0.5$. \textbf{Right:} The volume filling factor for fast, slow, and Alfv\'enic mode as a function of $M_A$. All simulations have $M_S\approx6.0$. Circular symbol represents a resolution of $792^3$, while inverted triangle symbol represents the cases of $512^3$.}
\end{figure*}
\begin{figure*}
	\centering
	\includegraphics[width=1.0\linewidth]{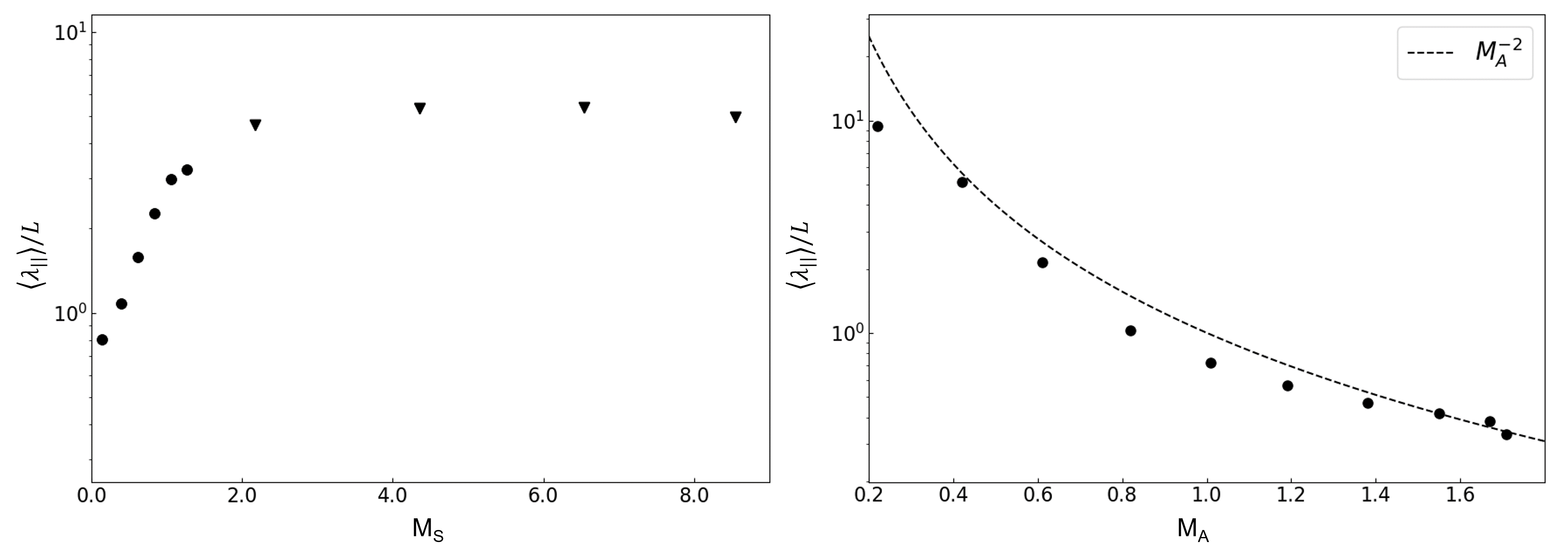}
	\caption{\label{fig:ms_lambda} \textbf{Left:} The mean free path $\lambda_\parallel$ in the unit of cube size $L$ as a function of $M_S$. All simulations have $M_A\approx0.5$. \textbf{Right:} The mean free path $\lambda_\parallel$ in the unit of cube size $L$ as a function of $M_A$. All simulations have $M_S\approx6.0$. Circular symbol represents a resolution of $792^3$, while inverted triangle symbol represents $512^3$.}
\end{figure*}
\section{Results}
\label{sec:results}
\subsection{Properties of MHD turbulence}
Fig.~\ref{fig:vspectrum} presents the kinetic energy spectrum. The velocity is decomposed into fast, slow, and Alfv\'enic modes using the decomposition method given in \S~\ref{sec:method}. Because magnetic field fluctuation is proportional to velocity fluctuation: $|b_k|=(B_0v_k/c_p)|\boldsymbol{B_0}\times\hat{\boldsymbol{\zeta}}|$, the spectrum also reflects magnetic field's properties. Here $b_k$ and $v_k$ are the magnetic field fluctuation and velocity fluctuation in Fourier space, respectively. $c_p$ denotes the propagation speed of the slow, fast, Alfv\'enic mode and $\hat{\boldsymbol{\zeta}}$ is the corresponding displacement vector. The slope of Alfv\'enic mode's spectrum is close to -5/3 as a result of the Kolmogorov scaling. The slope is insensitive to the value of $M_A$ and $M_S$, although the spectrum's amplitude increases for large $M_S$ cases due to more injected kinetic energy. As for fast mode, the slope gets steeper than $-5/2$ being close to the value of $-2$. $-2$ usually corresponds to shocked gas in supersonic condition and \citet{2003MNRAS.345..325C} found a slope of $-3/2$ for subsonic turbulence. The steeper slope indicates fast mode dissipates its energy quickly. However, a discrepancy of the slope has been reported by \citet{2010ApJ...720..742K} and \cite{2021arXiv210910977K}, who also found $k=-2$ for subsonic turbulence. Our results agree with the latter studies. In terms of the slow mode, it behaves in a way similar to Alfv\'en mode. Its slope is, therefore, still -5/3.

Additionally, to investigate the anisotropy of each mode, we employ the second-order structure function $SF(R,z)$ defined in the local reference frame, which is widely used to statistically describe a turbulent field;
\begin{equation}
\label{eq.st_loc}
\begin{aligned}
    &SF(R,Z)=\langle|\boldsymbol{v}(\boldsymbol{r_1})-\boldsymbol{v}(\boldsymbol{r_2})|^2 \rangle
\end{aligned}
\end{equation}
where $\boldsymbol{v}(\boldsymbol{r_1})$ and $\boldsymbol{v}(\boldsymbol{r_2})$ give the velocity at positions $\boldsymbol{r_1}$ and $\boldsymbol{r_2}$. We define the local magnetic field following \cite{2002ApJ...564..291C}:
\begin{equation}
\label{eq.st_loc}
\begin{aligned}
    &\boldsymbol{\Tilde{B}}=\frac{1}{2}(\boldsymbol{B}(\boldsymbol{r_1})-\boldsymbol{B}(\boldsymbol{r_2}))\\
\end{aligned}
\end{equation}
where $\boldsymbol{\Tilde{B}}$ is the local magnetic field defined by a cylindrical coordinate system $R$ and $Z$. In this system, axis are defined as  $R=|\hat{\boldsymbol{Z}}\times(\boldsymbol{r_1}-\boldsymbol{r_2})|$ and $Z=\hat{\boldsymbol{Z}}\cdot(\boldsymbol{r_1}-\boldsymbol{r_2})$ with $\hat{\boldsymbol{Z}}=\boldsymbol{\Tilde{B}}/|\boldsymbol{\Tilde{B}}|$.

As an example, the structure functions for simulation A1 ($M_S=0.62, M_A=0.56$) and C2 ($M_S=6.47, M_A=0.61$) are presented in Fig.~\ref{fig:SF_map}. The calculation randomly uses half million data points. We can see that in both subsonic and supersonic conditions, the iso-contours of Alfv\'enic mode and slow mode are elliptical. The semi-major axis is along the $Z$ direction, which is parallel to the local magnetic field. Therefore, the Alfv\'enic and slow mode are anisotropic. However, for fast mode, its iso-contours are more close to isotropic, as suggested by \citet{2003MNRAS.345..325C}. In Appendix~\ref{appendix.A}, we plot the structure function of velocity fluctuations exactly parallel and perpendicular to local magnetic fields, i.e., $SF(0,Z)$ and $SF(R,0)$, respectively. As expected for slow and Alfv\'en mode, the fluctuation is more significant in the direction perpendicular to the local magnetic fields. This minimum fluctuation, therefore, appears to the local magnetic field direction \citep{SFA}. As for fast mode, the perpendicular fluctuation is slightly higher than the parallel fluctuation. The difference, however, is insignificant and the fast mode is isotropic.

Also, we examine the relative significance of fast, slow, and Alfv\'enic modes in various astrophysical conditions, including subsonic warm gas and supersonic cold gas. We calculate the kinetic energy for each mode over the entire box as:
\begin{equation}
    E_{A,s,f}=\int_V v_{A,s,f}^2dV
\end{equation}
where the subscripts $A,s,f$ denote for Alfv\'enic, slow, and fast modes, respectively. The energy fraction for each mode is defined as its fraction in the total energy $E_{tot}=E_A+E_f+E_s$. The results are presented in Fig.~\ref{fig:mode}. Note that we only decompose sub-Alfv\'enic turbulence here, as super-Alfv\'enic case appears no mean magnetic field so that the decomposition method is not appropriated. We find that the fast mode always has a smaller fraction of energy than the other two modes. In terms of a fixed $M_A\approx0.5$, the fraction of fast mode decreases until $M_S\approx2$ but get saturated at the value of $\approx10\%$ when $M_S>2$. This trend exactly agrees with the change of mean free path (see Fig.~\ref{fig:ms_lambda}), as fast mode is the major agency of pitch angle scattering. Our results suggest that fast mode's fraction, as well as the mean free path, is insensitive to large $M_S$. In addition, the Alfv\'enic mode usually dominates over the other two modes in terms of energy fractions, occupying a fraction of $\approx50\%$ for supersonic turbulence. The trend we observe is consistent with the result from \cite{2011PhRvL.107k4504F}, who found that the fraction of solenoidal turbulence (i.e., the Alfv\'enic mode) is a nearly a constant irrespective of $M_S$ using a solenoidal driving. 

Another important factor, which can introduce more fraction of the fast mode, is $M_A$. Therefore, we fix $M_S\approx6$ and investigate the fraction of fast mode as a function of $M_A$ in Fig.~\ref{fig:mode}. We see that the fraction of fast mode increases when $M_A$ is large, while the fraction of Afv\'enic mode decreases. However, the fraction of slow mode stays at 40\% around, showing no apparent relation with respect to $M_A$.

\subsection{Parallel mean free path of test particles}

\begin{figure}
	\centering
	\includegraphics[width=1.0\linewidth]{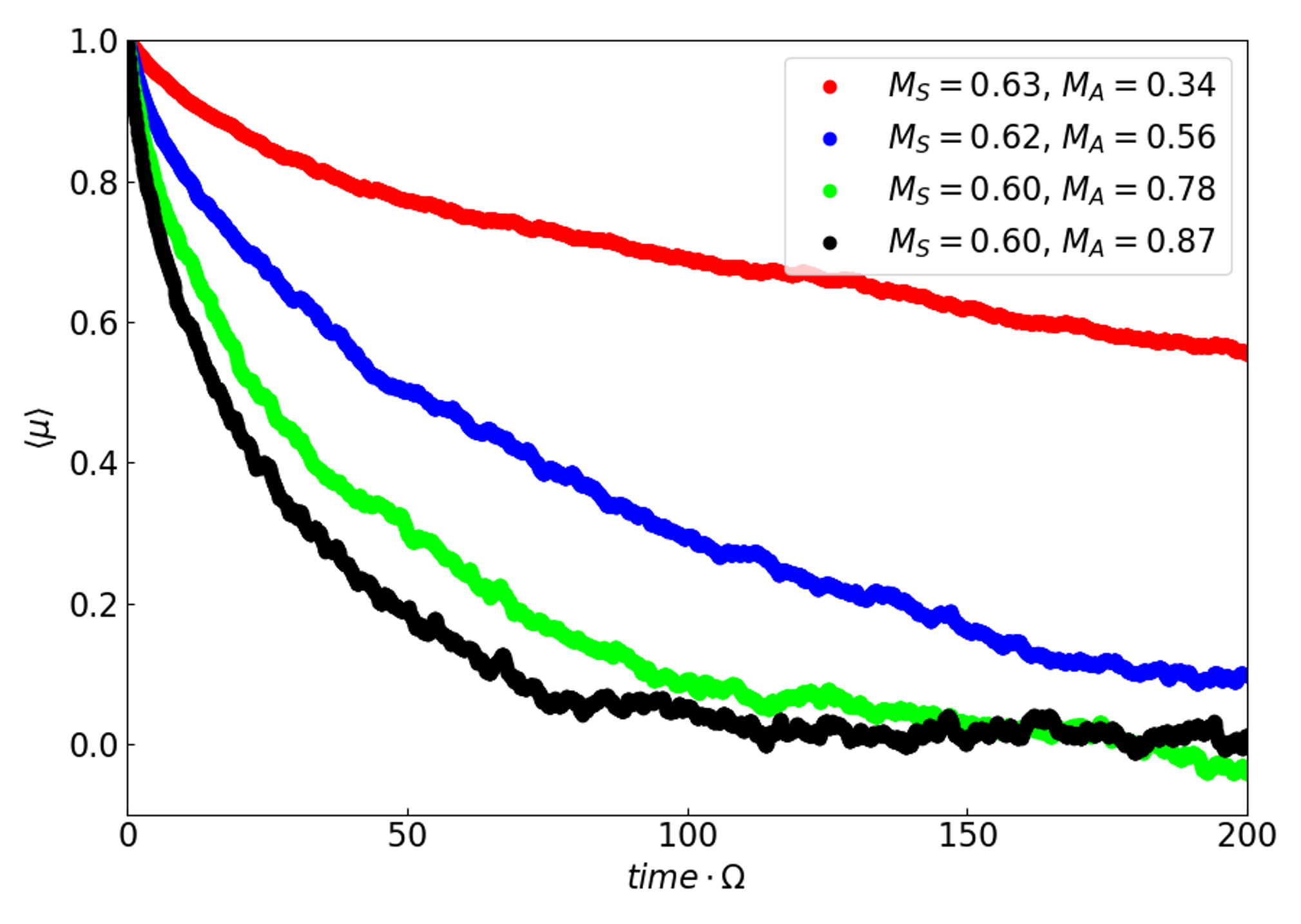}
	\caption{\label{fig:mu}The evolution of averaged $\langle\mu\rangle$ over all particles, i.e., the cosine of pitch angle with respect to the CRs' gyro periods . }
\end{figure}

\begin{figure*}
	\centering
	\includegraphics[width=1.0\linewidth]{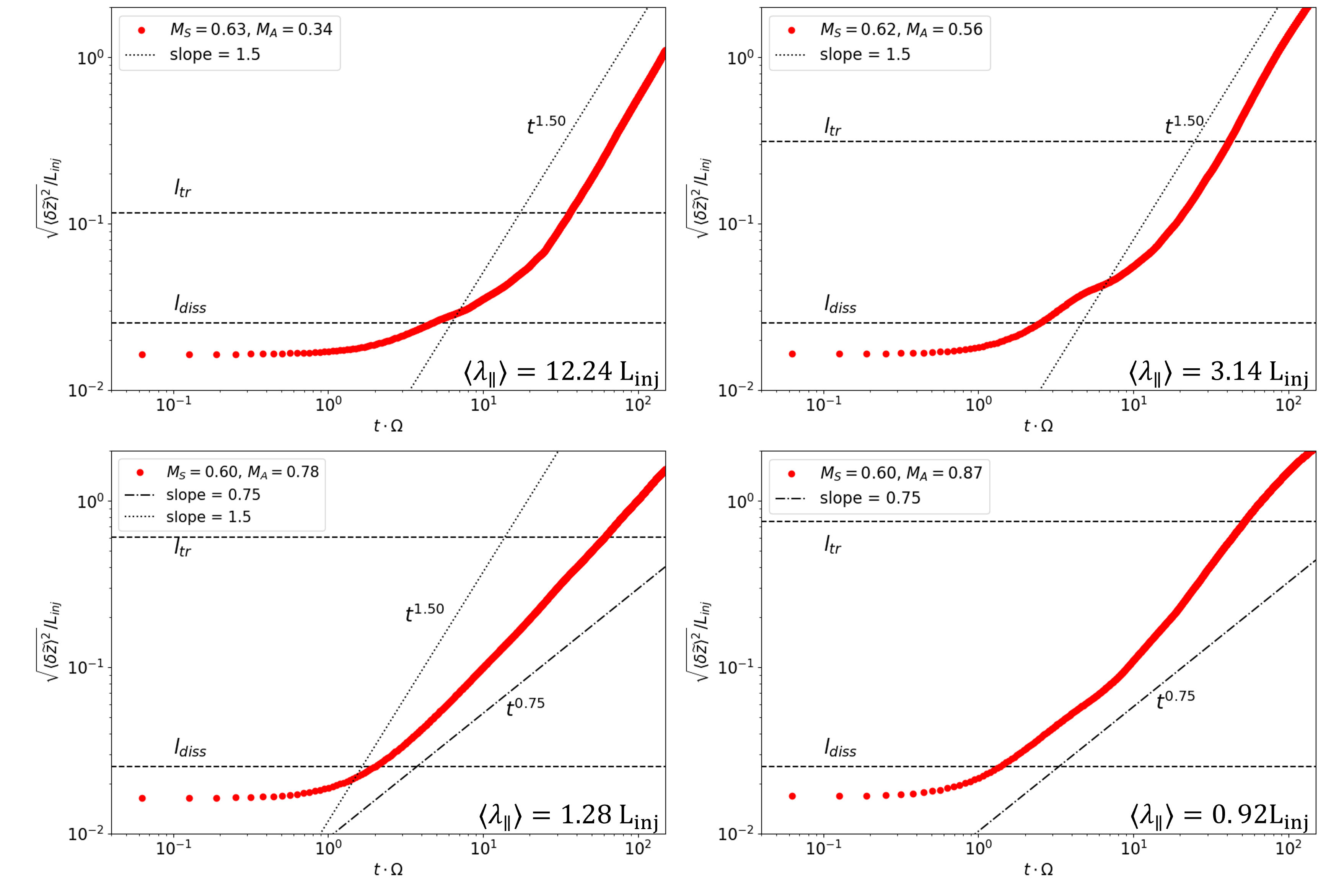}
	\caption{\label{fig:super_tot} The plot of perpendicular displacement of the particles $\sqrt{\langle\delta\Tilde{z}^2\rangle}$ versus the CRs' gyro periods $t\cdot\Omega$ in subsonic and sub-Alfv\'enic MHD turbulence. $l_{\rm tr}$ represents the transition scale, $l_{\rm diss}$ is the numerical dissipation scale, and $L_{\rm inj}$ is the injection scale.}
\end{figure*}
\begin{figure}
	\centering
	\includegraphics[width=1.0\linewidth]{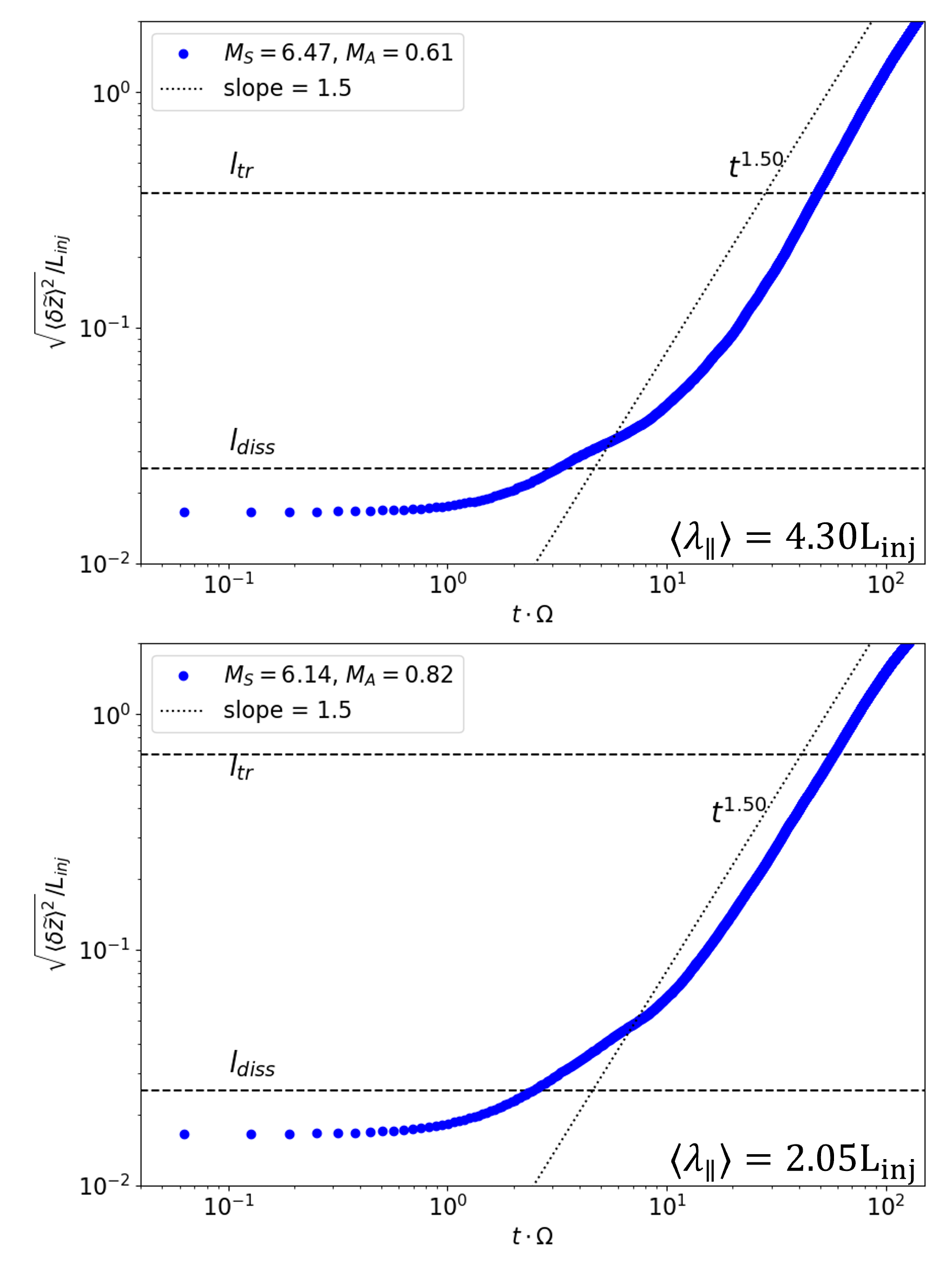}
	\caption{\label{fig:sub_run} Same as Fig.~\ref{fig:super_tot} but for sub-Alfv\'enic and supersonic ($M_S\approx6$) turbulence here.}
\end{figure}
\begin{figure}
	\centering
	\includegraphics[width=1.0\linewidth]{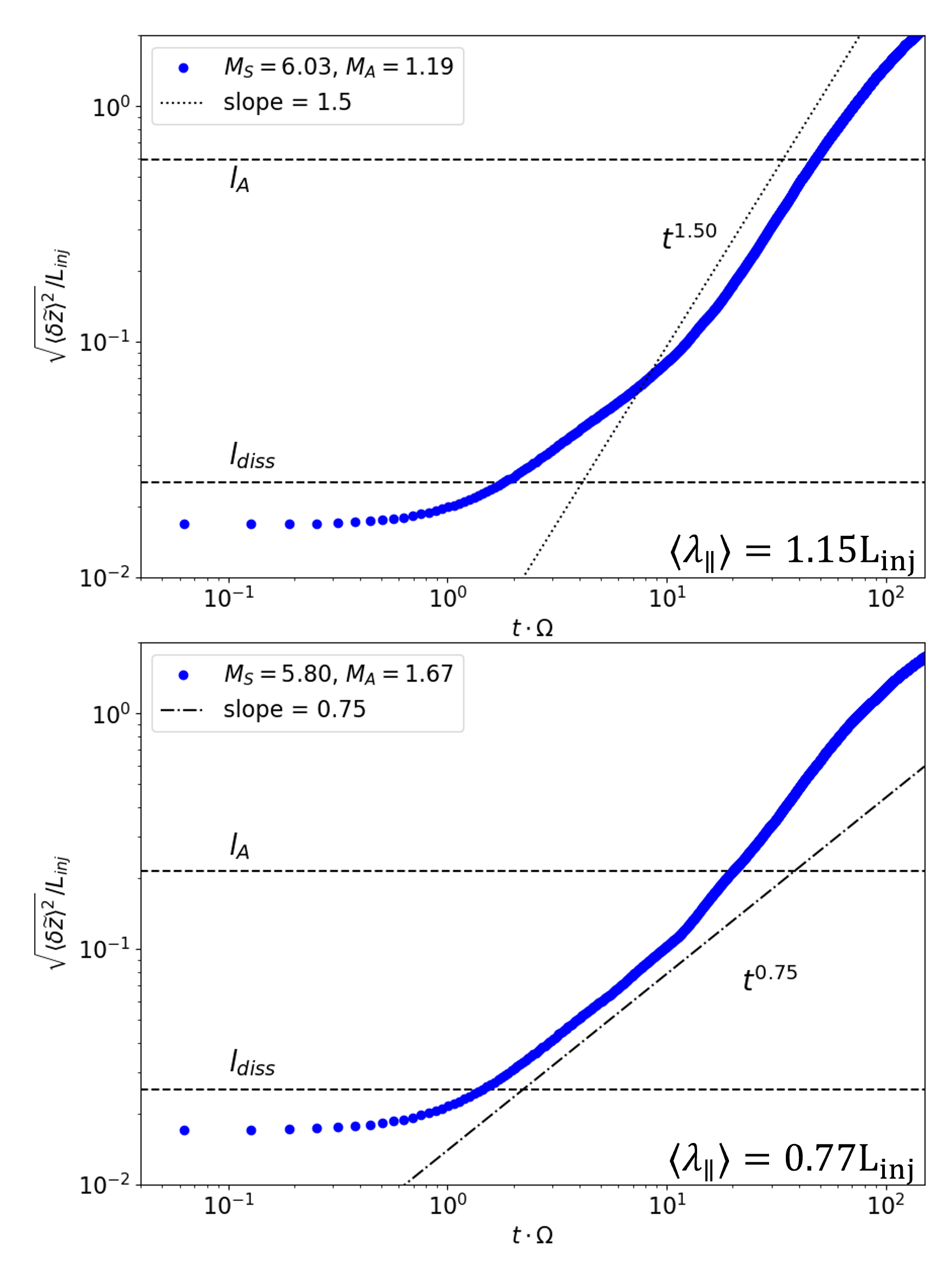}
	\caption{\label{fig:sup_run} The plot of perpendicular displacement of the particles $\sqrt{\langle\delta\Tilde{z}^2\rangle}$ versus the CRs' gyro periods
    $t\cdot\Omega$. $l_{A}=L_{\rm inj}M_A^{-3}$ represents the transition scale, $l_{\rm diss}$ is the numerical dissipation scale, and $L_{\rm inj}$ is the injection scale. We consider super-Alfv\'enic and supersonic ($M_S\approx6$) turbulence here.}
\end{figure}
To determine the mean free path of particles, we inject 1000 test particles, which are sufficient for the statistics \citep{XY13}, at random initial positions and initial pitch angles 0 degrees at a snapshot of the MHD simulation after the turbulent cascade is fully developed. The Larmor radius $r_L=u/\Omega$ is set as 1/50 of cube size $L$, where $\Omega=(qB)/(\gamma mc)$ is the frequency of a particle’s gyromotion ($\gamma$ is the particle’s Lorentz factor). We trace the change of pitch angle until it achieves 90 degrees. The corresponding averaged distance traveled along the local magnetic field is taken as the parallel  mean free path. 

In Fig.~\ref{fig:ms_lambda}, we present the mean free path $\langle\lambda_\parallel\rangle$ as functions of $M_S$ and $M_A$. We firstly fix $M_A\approx0.5$ and find $\langle\lambda_\parallel\rangle$ increases until $M_S\approx2$. The increment of $\langle\lambda_\parallel\rangle$ corresponds to weaker scattering, which indicates that the fast mode's fraction becomes less significant. However, when $M_S>2$, $\langle\lambda_\parallel\rangle$ stops increasing but being saturated at the value of 4 or 5 around and fast mode's fraction has little variation. In supersonic condition $M_S\approx6$, $\langle\lambda_\parallel\rangle$ decreases in a power-law manner as $M_A^{-2}$. As fast mode is the primary agent for scattering, a large $M_A$, therefore, corresponds to a small mean free path. It suggests that a observed small mean free path in supersonic condition is mainly contributed by relatively weak magnetic fluctuation, i.e., large $M_A$.

Fig.~\ref{fig:mu} shows the evolution of $\langle\mu\rangle$, i.e., the cosine of pitch angle as a function of time, averaged over 1000 particles. Obviously, the particles are scattered by magnetic fluctuations so that the averaged $\langle\mu\rangle$ monotonically decreases. We can also see that the decrease of $\langle\mu\rangle$ is slower for low $M_A$ cases, showing that the increased $M_A$ leads to enhanced efficiency in particle scattering. It can be understood as that the particles' gyroresonant scattering at a smaller $M_A$ is less efficient so that the corresponding mean free path is larger. 
Indeed, we find the mean free path parallel to the local magnetic field  $\langle\lambda_\parallel\rangle$ are $12.24L_{\rm inj}$, $3.14L_{\rm inj}$, $1.28L_{\rm inj}$, and $0.92L_{\rm inj}$ for the cases of $M_A=0.34$, $0.56$, $0.78$, and $0.87$, respectively. Here $L_{\rm inj}=0.5L$ is the injection scale.

\subsection{Perpendicular superdiffusion}
\subsubsection{Subsonic turbulence}
To test the perpendicular superdiffusion, we simultaneously inject 50 beams of test particles randomly in the simulation cube. Each beam contains 20 particles. The spatial separations between particles in each beam are $L/100$ pixels. Like the setting above, all the particles have $r_L = L/50$ and an initial pitch angle of 0. We trace the particle trajectories along the local magnetic fields at each time step. We also interpolate the trajectories so that we can measure the separation $\delta\Tilde{z}$ between the trajectories at each identical time step. The rms value $\sqrt{\langle\delta\Tilde{z}^2\rangle}$ is taken as the perpendicular displacement of the particles.

Fig.~\ref{fig:super_tot} presents the correlation of $\sqrt{\langle\delta\Tilde{z}^2\rangle}$ and the CRs' gyro periods $t\cdot\Omega$ in sub-Alfv\'enic and subsonic turbulence. Note that the superdiffusion only occurs within the inertial range of turbulence, which is half of the box size. The integration time is therefore determined not by the number of gyroperiods, but by the perpendicular displacement (see Appendix.~\ref{appendix.B}). We limit the gyro periods up to 200, in which the perpendicular separation has reached the box's size and the transport has reached an asymptotic regime. Due to the periodic boundary condition, the separation does not increase at a later time. We can see that the growth of $\sqrt{\langle\delta\Tilde{z}^2\rangle}$ is fast for large mean free path cases, as the scattering of particles is very inefficient. In contrast, small mean free path (i.e., $\langle \lambda_\parallel\rangle< L_{\rm inj}$) corresponds to a slowly increase of the perpendicular displacement. In particular, in the absence of scattering, i.e., $\langle \lambda_\parallel\rangle\gg L_{\rm inj}$, $\sqrt{\langle\delta\Tilde{z}^2\rangle}$ grows in a power-law relation with a power-law index $1.5$ after passing the numerical dissipation scale. The analysis for super-Alfv\'enic and subsonic turbulence was performed in \citet{XY13}. Nevertheless, when the role of scattering becomes more important, i.e., in the case of $\langle \lambda_\parallel\rangle= 1.28L_{\rm inj}$, the power-law index becomes shallower. This power-law index finally arrives at slope $=0.75$ when the scattering is significant, i.e., $\langle \lambda_\parallel\rangle<L_{\rm inj}$.

\subsubsection{Supersonic turbulence}
We repeat the analysis for highly supersonic turbulence in this section. Note because of the small energy fraction of fast mode, scattering is always weak so that $\langle \lambda_\parallel\rangle> L_{\rm inj}$ when $M_A\le0.8$. The result in sub-Alfv\'enic regime is presented in Fig.~\ref{fig:sub_run}. Like the subsonic cases above, when $\langle \lambda_\parallel\rangle\gg L_{\rm inj}$, $\sqrt{\langle\delta\Tilde{z}^2\rangle}$ grows among transport time $t$ in a power-law relation with a power-law index $1.5$ after passing the numerical dissipation scale. This superdiffusion of perpendicular displacement is still dominated by the diverging magnetic field lines in sub-Alfv\'enic and supersonic turbulence. 

In Fig.~\ref{fig:sup_run}, we further increase $M_A$ to super-Alfv\'enic regime. Similar to the subsonic cases (see Fig.~\ref{fig:super_tot}), scattering slows down the increment of perpendicular displacement $\sqrt{\langle\delta\Tilde{z}^2\rangle}$. When $\langle \lambda_\parallel\rangle> L_{\rm inj}$, $\sqrt{\langle\delta\Tilde{z}^2\rangle}$ grows among the magnetic field lines in a power-law relation $\sqrt{\langle\delta\Tilde{z}^2\rangle}\propto t^{1.5}$. In contrast, when $\langle \lambda_\parallel\rangle< L_{\rm inj}$, the scattering becomes efficient so that $\sqrt{\langle\delta\Tilde{z}^2\rangle}\propto t^{0.75}$.

\section{Discussion}
\label{sec.dis}

\subsection{Comparison with earlier studies}
Perpendicular propagation of CRs perpendicular magnetic fields has been a long-standing problem. While the cross-field diffusion due to resonance scattering is extremely slow, \cite{1969ApJ...155..777J} identified the magnetic field wandering as the dominant cause of CR displacement perpendicular to the galactic magnetic field. 

The quantitative study of the perpendicular propagation of CR became possible after the laws of magnetic wandering were identified in \citet{LV99}. This stimulated further quantitative research in the field. In the works that followed \cite{YL08, 2014ApJ...784...38L}, perpendicular diffusion was divided into superdiffusion into scales less than the turbulence injection scale\footnote{For super-Alfv\'enic turbulence one should use $l_A$ instead of the actual injection scale \citep{2014ApJ...784...38L}.} and normal diffusion on scales larger than turbulence injection scale. The former can be subdivided into two sub-cases, depending on whether the mean free path is larger or smaller than the turbulence injection scale. It was also predicted in \cite{2014ApJ...784...38L} that, depending on the relation the laws of the perpendicular propagation should change. As the mean free path $\langle\lambda_\|\rangle$ changes, the relation between the perpendicular displacements changes from $\sqrt{\langle z^2\rangle}\propto t^{3/2}$ to $\sqrt{\langle z^2\rangle}\propto t^{3/4}$. Both laws are superdiffusive, even though the rates of CRs dispersion growth are different.   

The pioneering studies numerically of CRs' (super)perpendicular and parallel diffusion include \cite{2006AGUSMSH42A..06Z, 2007PhPl...14a2311P,XY13}.
 \citet{XY13} focused on the study of superdiffusion in subsonic and sub-Alfv'enic turbulence with post-processed resonant slab fluctuations introduced in some cases. The study confirmed that both diffusive and superdiffusive propagation of CRs scale with $M_A^4$, which is different from the naive estimates in some accepted studies. However, \citet{2016A&A...588A..73C} reported the absence of superdiffusion in sub-Alfv\'enic case $M_A<0.7$, which maybe caused by limited range of strong MHD turbulence. A more recent numerical study in \citet{2021arXiv210801936M} is complimentary to our study of superdiffusion in the current paper. In particular, \citet{2021arXiv210801936M} perform the decomposition of MHD turbulence with various Alfv\'en Mach numbers and compare propagation both in the global and local reference frame. It provides a study of both CRs diffusion and superdiffusion for a fixed sonic Mach number. 
The study showed that the superdiffusion is better observed in the local magnetic reference frame for Alfv\'en mode in agreement with \citep{2014ApJ...784...38L}. 

In comparison with \citet{2021arXiv210801936M}, our paper is focused on numerical testing of the laws of superdiffusion for a variety of sonic Mach numbers $M_S$. We explore the interplay of $M_S$ and $M_A$ for the CR propagation in the variety of  turbulent regimes relevant to the multiphase ISM. To get better insight into CR propagation, we employ MHD simulations with higher numerical resolution. In particular, we investigate the significance of slow, fast, and Alfv\'en mode. We show that for a fixed $M_A$, the energy fraction of fast mode has little variation with the sonic Mach number for $M_S\ge2$, leading to the stabilization of the parallel mean free path of CRs.

\subsection{Importance of superdiffusion}

The superdiffusion of CRs at scales less than $L_{\rm inj}$ is an important regime since this condition holds in many astrophysical environments. For example, $L_{\rm inj}\approx100$ pc in ISMs \citep{2004ARA&A..42..211E} and $L_{\rm inj}\approx50$ pc in the galaxy M51 \citep{2011MNRAS.412.2396F}. For instance, the CRs' acceleration processes in shocks happen on scales comparable to or smaller (e.g,  $<0.1$ pc in the shock precursor; \citealt{2009ApJ...707.1541B,2017ApJ...850..126X}) than $L_{\rm inj}$. Therefore, the superdiffusion tested in this work can significantly affect some of the popular ideas about CR acceleration in shocks \citep{2014ApJ...784...38L,2020ApJ...903..105Z}. Naturally, the correct description of the CR propagation and related effects requires accounting for the superdiffusion. 

In addition, the effects of superdiffusion induce new elements of CR dynamics. For instance, (i) because spontaneous stochasticity of magnetic field lines, the effect retracing of CR is suppressed in turbulent environments \citep{YL08}. (ii) \cite{2021arXiv210608362L} find a new effect of CR parallel mirror diffusion in magnetic mirrors instead of CR trapping.

\subsection{Our advances}

This work numerically confirms the perpendicular supperdiffusion in various physical conditions, including sub-Alfv\'enic, super-Alfv\'nic, subsonic, and supersonic. We perform numerical study about three MHD modes, i.e., Alfv\'enic, fast, and slow mode. We find that for incompressible driving adopted, the energy fraction of fast mode, which is the main agent for scattering stop increasing for $M_S\ge2$. As a result, the mean free path or scattering is saturated at large $M_S$ (with a fixed $M_A$). Consequently, the scattering is inefficient, and the diverging magnetic field lines dominate the supperdiffusion at scales less than the injection scale $L_{\rm inj}$ in supersonic and sub-Alfv\'enic conditions. Nevertheless, we find the energy fraction of fast mode increases for super-Alfv\'nic driving, i.e., weak magnetic field, so that the scattering becomes significant. This finding is important in understanding the suppressed diffusion in supersonic molecular clouds \citep{2021ApJ...910..126S}, as suggested by recent gamma-ray observations. As the energy fraction of fast mode is insensitive to supersonic turbulence, it is likely that the suppressed diffusion is attributed to trans-Alfv\'enic or super-Alfv\'enic turbulence in molecular clouds, which is revealed in observation \citep{2016ApJ...832..143F,survey,2021arXiv210913670L}

\section{Summary} 
\label{sec:con}
The paper attempts to provide a realistic description of CRs' propagation in compressible MHD turbulence. This work comprehensively investigates the perpendicular superdiffusion across the mean magnetic field in different physical conditions, including sub-Alfv\'enic, super-Alfv\'nic, subsonic, and supersonic turbulence. We summarize our results as follows:
\begin{enumerate}
    \item We demonstrate that the diverging of magnetic field lines is dominated by Alfv\'enic turbulence and on the scales less than the the turbulence injection scale, which induces CR perpendicular superdiffusion.
    \item For a given magnetization level, i.e. for the fixed $M_A$, the energy fraction of fast mode in supersonic turbulence does not exceed $\approx0.2$, but is stabilized with the sonic Mach number for $M_S\ge2$, leading to the stabilization of the parallel mean free path of CRs. 
    \item For a given $M_S$, the energy fraction of fast mode increases in super-Alfv\'nic turbulence, while the fraction Alfv\'enic mode decreases. This results in the decreasing mean free path of CRs with increasing $M_A$.
    \item Our results suggest that the suppressed diffusion in supersonic molecular clouds may be attributed to trans-Alfv\'enic or super-Alfv\'enic turbulence.
    \item We confirm that when the parallel propagation scale $x$ is smaller than the injection scale, the scaling of the CRs' perpendicular displacement $\sqrt{\langle z^2\rangle}$ changes from $\propto t^{3/2}$ ($\langle\lambda_\parallel\rangle>x$) to $\propto t^{3/4}$ ($\langle\lambda_\parallel\rangle<x$). 
\end{enumerate}

\section*{Acknowledgements}
Y.H. acknowledges the support of the NASA TCAN 144AAG1967. A.L. acknowledges the support of the NSF grant AST 1715754 and NASA ATP AAH7546. S.X. acknowledges the support for this work provided by NASA through the NASA Hubble Fellowship grant \# HST-HF2-51473.001-A awarded by the Space Telescope Science Institute, which is operated by the Association of Universities for Research in Astronomy, Incorporated, under NASA contract NAS5- 26555. 

\section*{Data Availability}
The data underlying this article will be shared on reasonable
request to the corresponding author.





\appendix

\section{Structure function for fast, slow, and Alfv\'en mode}
\label{appendix.A}
\begin{figure*}
	\centering
	\includegraphics[width=1.0\linewidth]{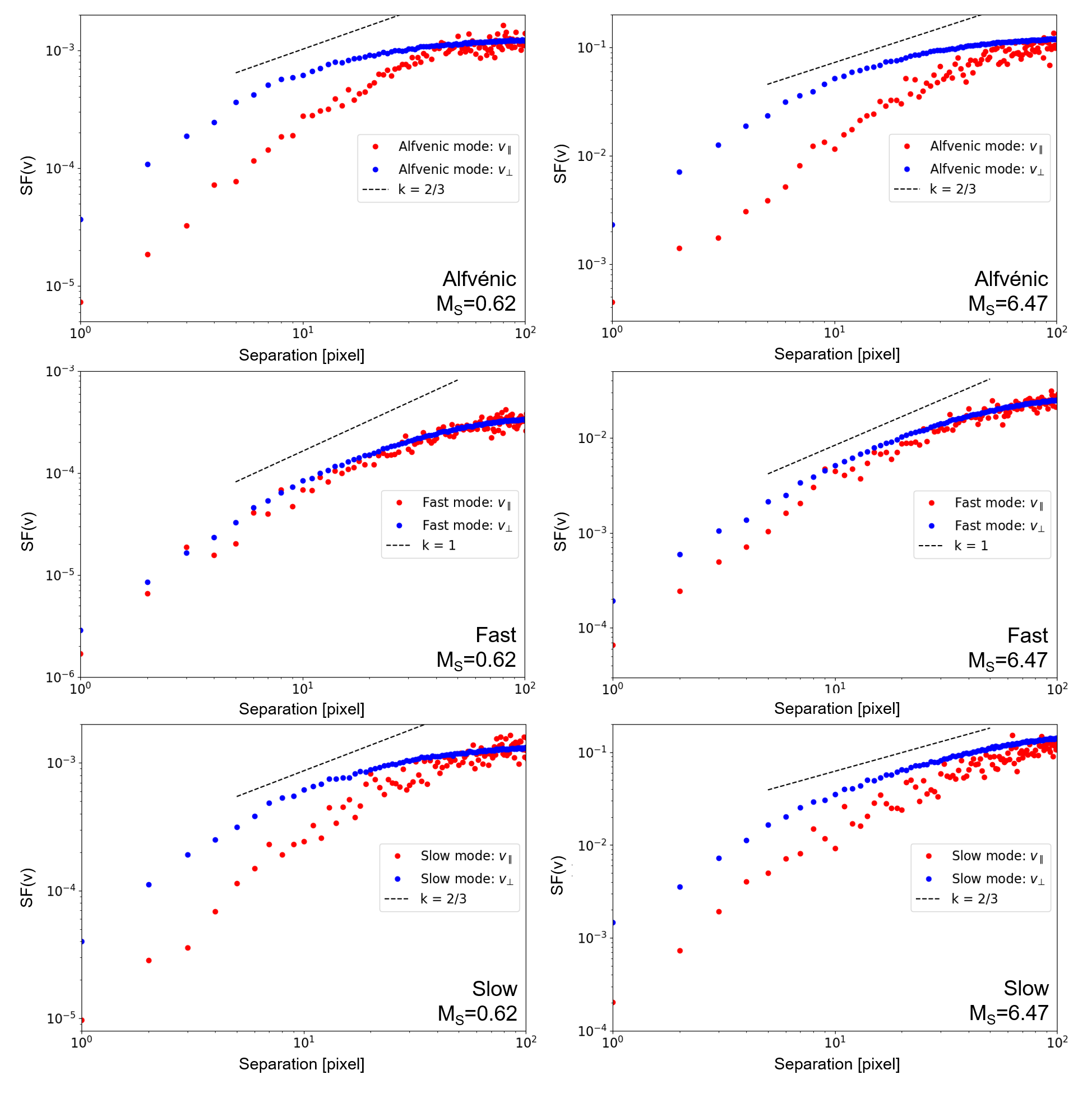}
	\caption{\label{fig:SF} \textbf{Left:} The structure-function of velocity fluctuations for Alfv\'enic (top), fast (central), and slow (bottom) mode, in the condition of $M_{\rm A}\approx0.5$. The fluctuations are decomposed into parallel (red) and perpendicular (blue) components with respect to local magnetic field. \textbf{Right:} Same as the left panel, but in supersonic regime. $k$ denotes the slope of reference line.}
\end{figure*}

\begin{figure*}
	\centering
	\includegraphics[width=1.0\linewidth]{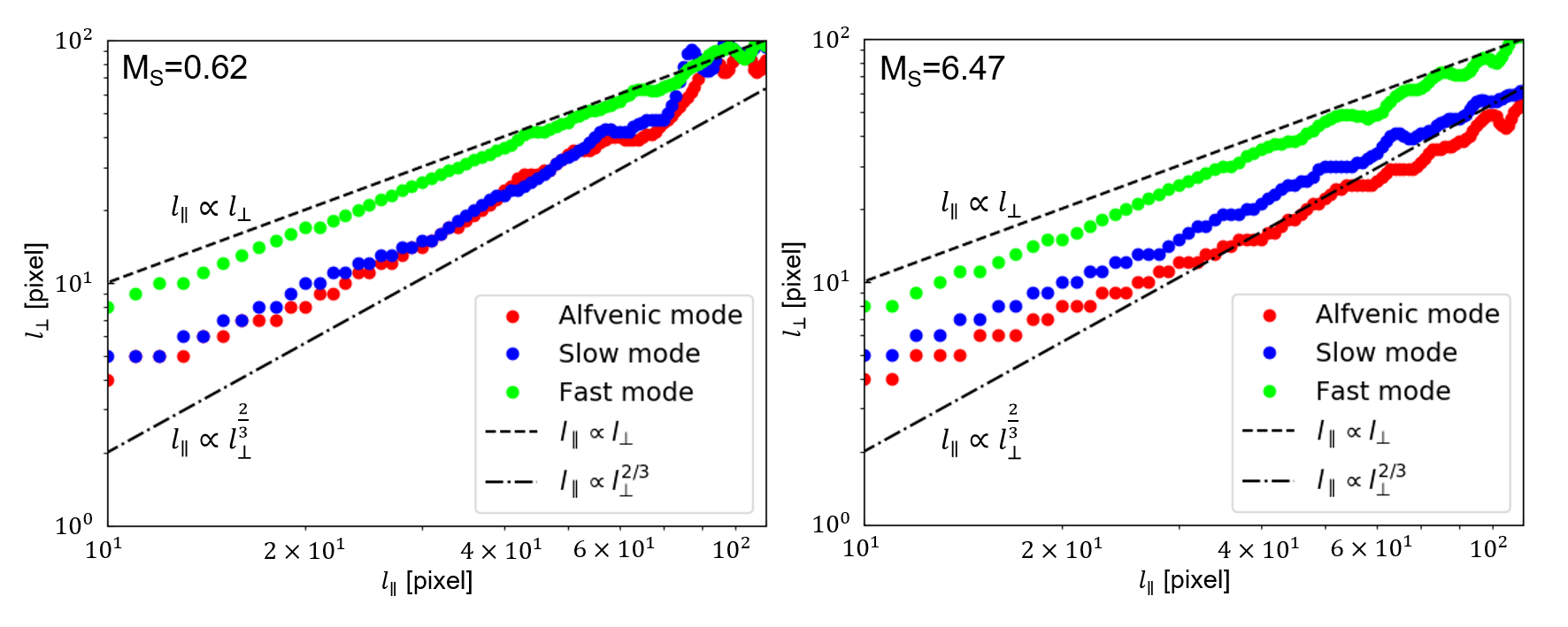}
	\caption{\label{fig:SF_ani} Anisotropy of the velocity's structures function for fast, slow, and Alfv\'en mode. To show anisotropy, we find the velocity iso-contour (see Fig.~\ref{fig:SF_map}). The values of $R$ and $Z$ (i.e., $SF(R,0)=SF(0,Z)$) at the same iso-contour are defined as $l_\bot=R$ and $l_\parallel=Z$, respectively. Both panels are drawn in the condition of $M_{\rm A}\approx0.5$.}
\end{figure*}
Here we plot the second-order structure function $SF(R,z)$ defined in the local reference frame:
\begin{equation}
\label{eq.st_loc}
\begin{aligned}
    &SF(R,Z)=\langle|\boldsymbol{v}(\boldsymbol{r_1})-\boldsymbol{v}(\boldsymbol{r_2})|^2 \rangle
\end{aligned}
\end{equation}
where $R$ and $Z$ define a cylindrical coordinate system for local magnetic field. 
In Fig.~\ref{fig:SF}, we plot the structure-function of velocity fluctuations exactly parallel, i.e., $SF(0,Z)$, and perpendicular, i.e., $SF(R,0)$ to local magnetic fields in subsonic and supersonic conditions. As slow and Alfv\'en mode are anisotropic, their fluctuations in the direction perpendicular to the local magnetic fields are more significant than the parallel one. The power-law slope is close to 2/3. As for isotropic fast mode, the parallel and perpendicular fluctuations are similar, while the perpendicular fluctuation is slightly higher. Its power-law slope is close to 1, corresponding to an energy spectrum's slope of -2.

To show the anisotropy scaling , we find the velocity iso-contour so that $SF(R,0)=SF(0,Z)$. The corresponding values of $R$ and $Z$ are defined as $l_\bot=R$ and $l_\parallel=Z$, respectively. In Fig.~\ref{fig:SF_ani}, we plot the relation of $l_\parallel$ and $l_\bot$. Apparently, we see that fast mode is isotropic so that $l_\parallel\propto l_\bot$. However, slow mode and Aflv\'enic mode are anisotropic exhibiting the scaling $l_\parallel\propto l_\bot^{2/3}$. 

\section{Effect of different gyro periods}
\label{appendix.B}
\begin{figure*}
	\centering
	\includegraphics[width=1.0\linewidth]{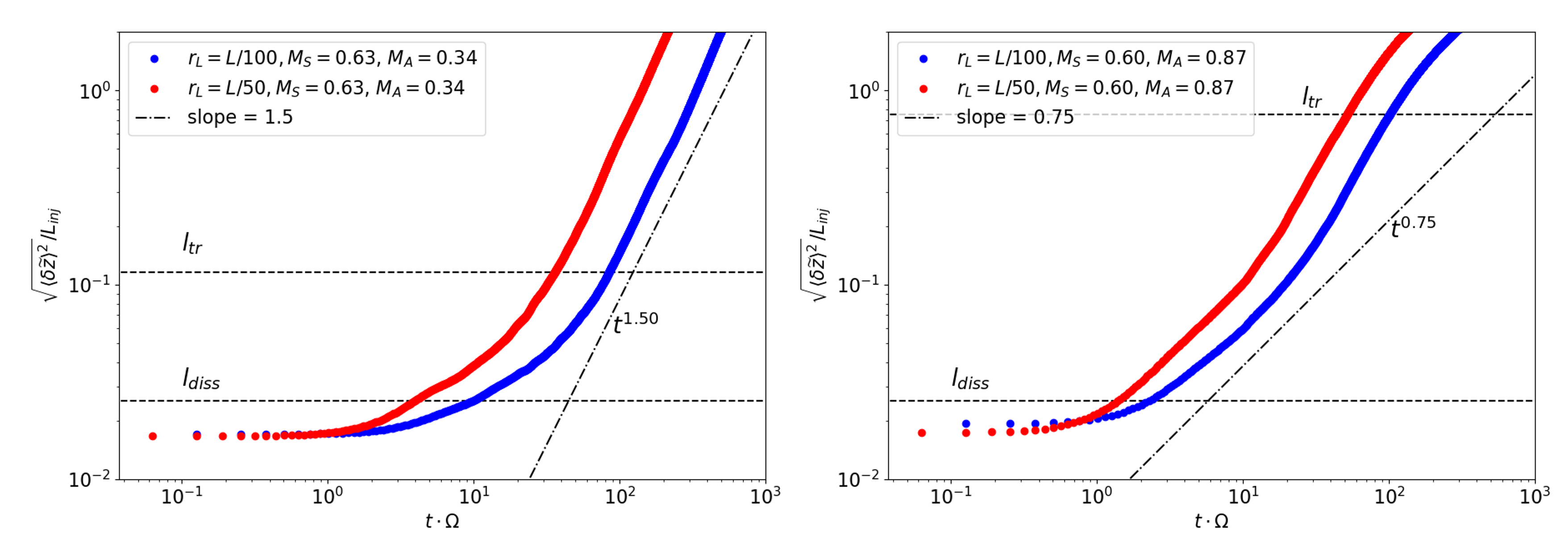}
	\caption{\label{fig:rl}The plot of perpendicular displacement of the particles $\sqrt{\langle\delta\Tilde{z}^2\rangle}$ versus the CRs' gyro periods $t\cdot\Omega$ in subsonic MHD turbulence. The gyro radius $r_L=L/100$ (blue) and $r_L=L/50$ (red) are compared for each condition. $l_{\rm tr}$ represents the transition scale, $l_{\rm diss}$ is the numerical dissipation scale, and $L_{\rm inj}$ is the injection scale.}
\end{figure*}

Fig.~\ref{fig:rl} presents the perpendicular displacement of the particles $\sqrt{\langle\delta\Tilde{z}^2\rangle}$ as a function the CRs' gyro periods $t\cdot\Omega$. The numerical setup is the same as the one used in \S~\ref{sec:results}, but here we use two values of gyro radius $r_L=L/100$ and $r_L=L/50$ separately. A small $r_L$ also indicates a small gyro period. Consequently, one needs a longer integration time to achieve the same perpendicular displacement value as large $r_L$. We can see that the superdiffusive behavior of perpendicular displacement is independent of $r_L$ and the integration time. The superdiffusion is determined by the CRs' displacement within the inertial range of turbulence. 

\bsp	
\label{lastpage}
\end{document}